\documentclass[journal,twoside]{IEEEtran} 
      \usepackage[utf8]{inputenc}
      \usepackage[T1]{fontenc}
\usepackage{amsmath,amsfonts}
\usepackage{algorithm}
\usepackage{array}
\usepackage[caption=false,font=normalsize,labelfont=sf,textfont=sf]{subfig}
\usepackage{textcomp}
\usepackage{stfloats}
\usepackage{url}
\usepackage{verbatim}
\usepackage{graphicx}
\graphicspath{ {./figures/} }
\usepackage{cite}
\usepackage{tikz}
\newcommand\copyrighttext{%
  \normalsize This work has been submitted to the IEEE for possible publication. Copyright may be transferred without notice, after which this version may no longer be accessible.
  }
\newcommand\copyrightnotice{%
\begin{tikzpicture}[remember picture,overlay]
\node[anchor=south] at (current page.south) {\fbox{\parbox{\dimexpr\textwidth-\fboxsep-\fboxrule\relax}{\copyrighttext}}};
\end{tikzpicture}%
}

\usepackage[noend]{algpseudocode}
\usepackage{array}
\usepackage{amssymb} 
\usepackage{amsthm}    
\usepackage{bm}
\usepackage{url}
\usepackage{xcolor}
\usepackage{soul}
\usepackage{enumitem}

\newtheorem{proposition}{Proposition}{\bfseries}{\itshape}
\newcommand\axiom[2] {\medskip\noindent{\bfseries #1.\ } {\rm #2}\par\medskip}

\newcommand{\diag}{\mathop{\mathrm{diag}}}

\def\sign{\mathop{\rm sign}\nolimits}
\newcommand{\x}[1]{} 
\newcommand\jJ[1]{{\footnotesize #1}}          
\newcommand\jI[1]{\!\!\jJ{#1}}          
\def\xy{\hspace*{.07em}}          
\def\xz{\hspace*{-.07em}}         
\def\paa#1{(\xz{\em #1\/})}

\def\cdc{,\ldots,}
\def\N{{\mathbb N}}

\def\F{{\cal F}}
\def\Ff{{\F^{40}}}
\def\G{{\cal G}}
\def\ve{\varepsilon} 

\def\1n{1\cdc n}
\newcommand{\eq}[2]{\begin{equation}\label{#1}#2\end{equation}}
\def\eqs*#1{\begin{eqnarray*}#1\end{eqnarray*}}

\def\beq#1{\begin{equation}\label{#1}}
\def\eeq{\end{equation}}

\def\Comm{\operatorname{Comm}}
\def\Katz{{\operatorname{Walk}}} 

\def\For{\operatorname{For}} 
\def\Hea{\operatorname{Heat}}
\def\Up#1{\vspace*{-#1em}}                                       
\def\BiPe{\mbox{\scriptsize(Bischeri,Peruzzi)}G_{15}}

\def\ToTo{\Leftrightarrow}
\def\ssm{\smallsetminus}

\def\Closeness{{\it Closeness\/}}

\def\WeightedDegree{{\it Weighted degree\/}}

\def\EstradaSubgraph{{\it Estrada subgraph\/}}
\def\Walk{{\it Walk\/}}
\def\Katzc{{\it Walk\/}} 
\def\KatzcCentrality{{\it Katz centrality\/}} 
\def\Forest{{\it Forest\/}}

\def\Communicability{{\it Communicability\/}}

\def\TotalCommunicability{{\it Total communicability\/}}
\def\TotalWalk{{\it Total walk\/}}
\def\Eccentricity{{\it Eccentricity\/}}
\def\ClosF{Closeness$($Forest$)$}
\def\ClosH{Closeness$($Heat$)$}
\def\ClolF{Closeness$(\log$Forest$)$}
\def\ClolFa{Closeness$^*(\log$Forest$)$}
\def\ClolH{Closeness$(\log$Heat$)$}
\def\ClolW{Closeness$(\log$Walk$)$}
\def\ClolWa{Closeness$^*(\log$Walk$)$}
\def\ClolC{Closeness$(\log$Communicability$)$}
\def\EcceF{Eccentricity$($Forest$)$}
\def\EcceH{Eccentricity$($Heat$)$}
\def\EcclF{Eccentricity$(\log$Forest$)$}
\def\EcclH{Eccentricity$(\log$Heat$)$}
\def\EcclW{Eccentricity$(\log$Walk$)$}
\def\EcclC{Eccentricity$(\log$Communicability$)$}
\def\separat{separat}
\definecolor{ggreen}{rgb}{0,.47,0}
\newcommand{\aB}[1]{#1}    
\newcommand{\aBa}[1]{#1}    

\setlist[enumerate]{topsep=.5em} 

\def\hl#1{#1}
\def\hla#1{#1}
\def\hlb#1{#1}
\begin{document}

\title{How to Choose the Most Appropriate Centrality Measure? A Decision Tree Approach}

\author{Pavel Chebotarev and Dmitry Gubanov
\thanks{The work of P.~Chebotarev was supported by the European Union (ERC, GENERALIZATION, 101039692). 
}
\thanks{P.~Chebotarev is with the Technion--Israel Institute of Technology, Haifa, 3200003 Israel 
\x{and the A.A.~Kharkevich Institute for Information Transmission Problems, RAS, 19 Bol'shoi Karetnyi per., Moscow, 127051 Russia }(e-mail: pavel4e@technion.ac.il\x{gmail.com}).}
\thanks{D.~Gubanov is with the Institute of Control Sciences, RAS, 65 Profsoyuznaya str., Moscow, 117997 Russia (e-mail: dmitry.a.g@gmail.com).}}
\markboth{IEEE Transactions on XXX, Vol. XX, No. XX, XXXX 2024}%
{P.~Chebotarev, D.~Gubanov\x{\MakeLowercase{\textit{et al.}}}: How to Choose the Most Appropriate Centrality Measure? A Decision Tree Approach}
\IEEEpubid{0000--0000/00\$00.00~\copyright~2024 IEEE}

\setcounter{page}{0}
\copyrightnotice

\maketitle
\begin{abstract}
Centrality metrics play a crucial role in network analysis, while the choice of specific measures significantly influences the accuracy of conclusions as each measure represents a unique concept of node importance. Among over 400 proposed indices, selecting the most suitable ones for specific applications remains a challenge. Existing approaches---model-based, data-driven, and axiomatic---have limitations, requiring association with models, training datasets, or restrictive axioms for each specific application. To address this, we introduce the culling method, which relies on the expert concept of centrality behavior on\x{ small} simple graphs. The culling method involves forming a set of candidate measures, generating a list of as small graphs as possible needed to distinguish the measures from each other, constructing a decision-tree survey, and identifying the measure consistent with the expert’s concept. We apply this approach to a diverse set of 40 centralities, including novel kernel-based indices, and combine it with the axiomatic approach. Remarkably, only 13 small 1-trees are sufficient to separate all 40 measures, even for pairs of closely related ones. By adopting simple ordinal\x{qualitative} axioms like Self-consistency or Bridge axiom, the set of measures can be drastically reduced making the culling survey short. Applying the culling method provides insightful findings on some centrality indices, such as PageRank, Bridging, and dissimilarity-based Eigencentrality measures, among others. The proposed approach offers a cost-effective solution in terms of labor and time, complementing existing methods for measure selection, and providing deeper insights into the underlying mechanisms of centrality measures.
\end{abstract}

\section{Introduction}
\label{s:intro}

\IEEEPARstart{U}{nderstanding} networks relies heavily on centrality metrics. However, the vast number of proposed point centrality measures, which now exceeds 400 and continues to grow \cite{centiserver} (cf.~\cite{Lu16Vital,Saxena20CentrSurvey-,Wan21Survey}), poses a challenge when it comes to selecting the most appropriate centrality indices for specific applications.

In some cases, researchers have a mathematical model that accurately captures the influence of nodes in the process of interest, naturally suggesting a measure of node importance that can be interpreted as centrality \cite{friedkin1991theoretical,borgatti2005centrality,wang2010electrical,chkhartishvili2018social,gubanov2018influence,Tyloo19oscill,Saxena20Models-}. 

However, in many cases, no detailed model is available, while the centrality of network nodes needs to be estimated.\x{there is a need to measure centrality of network nodes.} 
To address this, researchers rely on studying the typology of the application and theoretical taxonomy of centrality indices
\cite{borgatti2005centrality,Lu16Vital,Saxena20CentrSurvey-,Vignery20Methodology,Nooraie20Fusion,Wan21Survey,BlochJackson23}
to narrow down the options.
Further, specific measures can be compared experimentally by studying correlations, regressions, and appropriate distances between them and\x{ ``ground truth,'' i.e.,}
external network characteristics \cite{bolland1988sorting,rothenberg1995choosing,batool2014towards,Lu16Vital,Zhao16Competitiveness,Kandhway16diffus,%
oldham2019consistency,CaoBu19Smart,Rastogi20Correlative,Brysbaert21Protein,Hu22Influen,BlochJackson23}
and (or) using Principal Components Analysis (PCA) \cite{Vignery20Methodology,Sakellariou21Neurocraft}. 
\x{However}On the other hand, such studies yield varying results across datasets within the same subject area \cite{ashtiani2018systematic}
and do not uncover the underlying causes of correlations.

Hence, to supplement the experimental perspective, it is essential to focus on the\x{ intrinsic} properties of centralities\x{y measures} and the conditions they satisfy.
The quintessence of this is the axiomatic approach, which seeks to characterize\x{ assigns to} a measure by a minimal set of axioms\x{ that} it uniquely satisfies
(we refer to~\cite{holzman1990axiomatic,vohra1996axiomatic,monsuur2004centers,garg2009axiomatic\x{,kitti2016axioms},
dequiedt2017local,skibski2018axioms,WasSkibski23Pagerank\x{,KosnySkibski21fourmeasures}} for several examples). 
In \cite{BlochJackson23}, typology is incorporated into this\x{e axiomatic} approach\x{ through} by the use of nodal statistics.

However, the axiomatic approach has its limitations. Some of them are as follows.

1. 
Only a minority of measures are axiomatized, since constructing a suitable set of axioms is a non-trivial task.

2. For parametric families of measures 
(such as those in \cite{agneessens2017geodesic,csato2017measuring,deAndrade2019p}),
selecting the parameter value rarely has an axiomatic solution.

3. In many axiomatics, there is at least one technical axiom, which is not appealing\x{attractive} in itself and rather specifies the functional form of a centrality\x{ measure} (see, e.g., Closeness, Degree, and Decay axioms in \cite{garg2009axiomatic}, Linearity in \cite{dequiedt2017local}, Neighbor separability in \cite{Mago18Power}). It can be argued that adopting such an axiom does not fundamentally differ from adopting a\x{ centrality} measure\x{ itself} at once.
\IEEEpubidadjcol

4. Some non-technical axioms have rather sophisticated formulations, which makes it difficult to assess their desirability.

5. 
Comparing multiple axiomatics can be laborious, with no guarantee of reaching a definitive choice among them.

6. 
The rankings of nodes in simple networks provided by a\x{ unique} measure satisfying a set of axioms may be counterintuitive to users, despite the initial appeal\x{ of most} of the axioms.

For example, the \WeightedDegree\ {\em centrality}\x{c} \cite{bandyopadhyay2017generic-}
$$
f(u)=\sum_{v\in V,\,v\ne u}\frac{D(v)}{d(u,v)},
$$
where $V$ is the node set of a graph $G,$ 
$D(v)$ is the degree of node $v$, $d(u,v)$ being the shortest path distance between $u$ and $v,$ implements the reasonable idea of measuring the centrality of $u$ by the sum of the degrees of all other nodes with weights decreasing with distance from~$u.$
This measure satisfies four out of six axioms considered in \cite{bandyopadhyay2017generic-} (while Betweenness\x{c} or Eigenvector centrality\x{c} satisfy three axioms and no measure under study\x{ consideration} satisfies five or six) and its characterization is likely to be\x{ completed} obtained by adding a suitable
\x{arithmetically binding (algebraically restrictive)}axiom of an algebraic nature. However\x{At the same time}, for any {\em star\/} (a graph with $n>2$ nodes and $n-1$ edges incident to the same node called the center)\x{ with $n>2$ nodes}, this index considers the center to be less central than the leaves. \x{This (means suggests) leads to the conclusions}It follows that $(i)$~the Weighted degree\x{c} measures something other than centrality; $(ii)$~\x{in some cases, }the performance of a measure on simple examples\x{ can say more about it} is no less informative than the underlying heuristics or axioms it satisfies.\x{ it is dangerous to chose a centrality measure based on heuristics and the {\em number\/} of conditions it meets.}

The main observation that prompted\x{ gave rise to, served as the starting point for} this study was that \x{people}professionals familiar with a particular application often rely on their experience (frequently based on objective results) to compare node centrality in simple networks.
In this case, we can\x{ try to} select\x{offer (him/her) them} a centrality measure that ranks the nodes in the same way\x{similarly}. 
\x{In some cases, it}Furthermore, such a centrality can be \x{searched}sought in the set of measures that satisfy the\x{ normative conditions} axioms that the expert considers\x{ the most} important.

\begin{figure*}[!t] 
\begin{center}
\includegraphics[trim={0.0cm 5.98cm 0 5.7cm},clip,width=14.6cm]{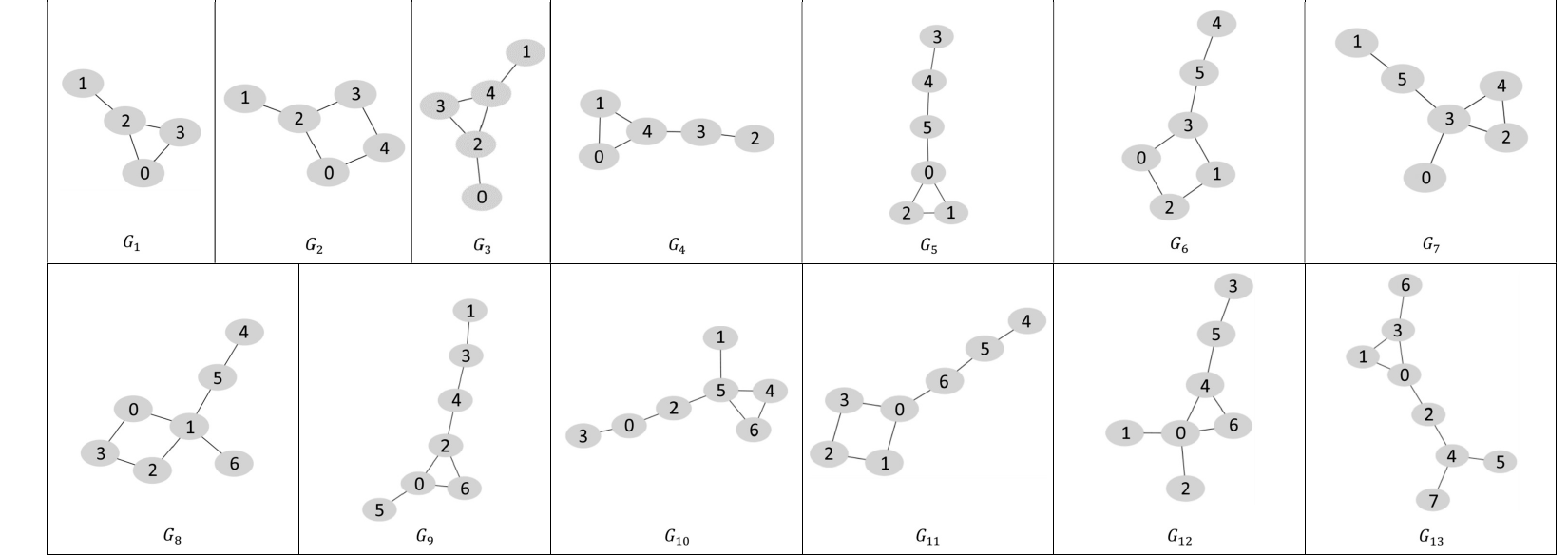} 
\\\vspace{-0.3em} 
\caption{Distinguishing graphs for \x{a}the test set $\Ff$ of 40 centrality measures.\label{f:SepGraphs}}
\end{center}
\end{figure*}

As another instructive example, consider\x{ one of the most famous centrality indices,} the PageRank centrality \cite{BrinPage1998anatomy,langville2006google\x{,newman2018networks},Avrachenkov22Book}, which was originally proposed for directed\x{ influence} reference networks (see \cite{FriedkinJo14PR,friedkin1991theoretical}) and has\x{ several} axiomatizations \cite{\x{palacios2004measurement,}altman2005ranking,WasSkibski23Pagerank\x{,KosnySkibski21fourmeasures}}. \x{Observe}Note that it sets $f(4)>f(5)$ (node 4 is more central than node~5) for the graph $G_5$ in Fig.~\ref{f:SepGraphs}. As well as some other inequalities it produces\x{ judgements it makes, results it provides} (namely, $f(4)>f(0)$ for~$G_2$; $f(1)>f(0)$ for~$G_7$; $f(0)>f(2)$ and $f(1)>f(5)$ for~$G_9$), this seems rather counterintuitive.
On the other hand\x{ However}, an analysis of the mechanism of this centrality reveals \cite{Che23JCNCentr} some types of real-world undirected networks that may require such a measure.
This \x{type }kind of singularity of measures may\x{are likely to} go unnoticed in an axiomatic study, but it is\x{ very} easy to detect when\x{ if we test} centralities are tested on a set of specially selected graphs.

Taking into account the above limitations of the axiomatic strategy, in this paper we \x{develop}propose an alternative approach based on expert beliefs\x{ the expert's opinion} on how the desired\x{required} centrality\x{ measure} should act on\x{ a set of} test\x{simple} graphs.

\subsection*{Purpose of the Study} 
\label{ss:Select}

\noindent We consider centrality measures on undirected graphs.
Given\x{ such} a graph $G=(V,E)$ with node set $V=V(G)$ such that $|V(G)|>3$ and edge set $E=E(G)$, a \emph{centrality measure\/} $f$ associates a real number $f(v)$ to each node $v\in V(G)$. This number is interpreted as follows\x{ the \emph{centrality\/} of $v$ in $G$ assigned by~$f$}: the larger $f(v),$ the more central $v$ is considered.
While $f(v)$ depends on $G,$ for simplicity we do not reflect this in the notation as only one graph is considered at a time.
We confine ourselves\x{restrict our consideration} to connected graphs, since there are centralities defined only for them.
Let $\F=\{f_1\cdc f_m\}$\x{, $m>1,$} be a set of centrality measures, and each $f_i$ is applicable to any graph of the above type.

In this paper, we are mainly interested in the \emph{order\/} of centrality values, not the numerical values themselves.
Let us say that a graph $G$ {\it distinguishes $($\emph{or} separates$)$
measures $f_i$ and $f_j$ on a pair of nodes $\{u,v\}$ $(u,v\in V(G))$} whenever\x{ if and only if}
\beq{e_disti} 
\sign(f_i(u)-f_i(v))\ne\sign(f_j(u)-f_j(v)).
\eeq

In words \eqref{e_disti} means that $f_i$ and $f_j$ disagree in comparing the centrality of $u$ and~$v.$

Centrality measures $f_i$ and $f_j$ are \emph{rank equivalent\/} if and only if there exists\x{is} no graph\x{ $G$} that distinguishes them\x{ $f_i$ and~$f_j$}.
We suppose that no distinct measures in $\F$ are rank equivalent.
Eliminating rank equivalent ``twins'' of some measures can be done by considering a large enough series of large enough graphs: if no graph distinguishes a couple of measures,\x{ give the same centrality ranking on each graph,} we remove one of them, even if we have not been able to rigorously prove that they are rank equivalent.

The\x{ purpose(goal)} aim of this\x{ paper} work is to propose\x{develop} a\x{ tool} method that allows the expert\x{user} to select\x{choose} a centrality measure that matches\x{based on} their application-specific understanding of how such a measure should act on a\x{ set} collection of test graphs.

\x{To do so}Given a set of measures $\F$, we need to construct:

\begin{itemize}[itemsep=-.10em,topsep=.0em]
\item
  a collection of fairly simple graphs that contains\x{includes} a distinguishing graph for every pair of distinct measures $f_i, f_j\in\F;$
\item
  a questionnaire such that every\x{ series} set of responses to its questions determines a unique measure.
\end{itemize}

Ideally, these collection and questionnaire should be generated and modified automatically whenever a new set $\F$ is provided or an existing set $\F$ is updated.

The method that consists in constructing such collections and questionnaires and passing the corresponding surveys to select the most appropriate centrality measures will be called \emph{culling}. 
We develop\x{introduce describe} this method in the following section. The computer code that implements it is provided at {\small\url{https://github.com/DAGubanov/centralities_dec_tree}}.

\section{Results}

\noindent We now present the main procedures of the culling method and then illustrate its application.

\subsection{Generating Graphs that Distinguish Measures}
\label{ss:Genera}

\noindent Let a set $\F$ of rank-nonequivalent centrality measures be given. 
We aim to construct a\x{ sequence} list $\G=(G_1\cdc G_K)$ of graphs that \separat{e} all measures in~$\F.$ 

The distinguishing graphs should be simple enough for experts to answer which \hla{pairwise centrality order}, i.e., $f(u)>f(v)$ or $f(u)<f(v)$ or $f(u)=f(v)$ is most consistent with their application-specific understanding 
based on their experience and previous studies.

For our experiments\x{ first surveys presented in Section~ \ref{ss:TestSu}}, we opted for the class of \emph{unicyclic graphs\/} (also called unicycles, $1$-trees, or augmented trees). These are connected graphs with the number of edges equal to the number of nodes. Such a graph can be obtained from a tree by connecting any two non-adjacent nodes with an edge.

The advantage of $1$-trees is that they are\x{ very} simple and\x{ perhaps} can be even more comfortable to perceive and compare the centrality of nodes than trees, thanks to an eye-catching cycle (\x{see }Fig.~\ref{f:SepGraphs}).

The procedure of constructing the list $\G=(G_1\cdc G_K)$ of \separat{ing} $1$-trees is as follows.

\begin{enumerate}[itemsep=-.1em,topsep=-.1em]
\item[1.]
At the beginning, the list $\G$ is empty.
\item[2.]
We sequentially generate unlabeled trees, starting from the trees with\x{ three} four nodes and increasing the number of nodes.
\begin{enumerate}[itemsep=-.1em,topsep=-.2em]
\item[2.1.]
For each tree, we produce various $1$-trees by adding different edges that connect non-adjacent nodes.
\begin{enumerate}[itemsep=-.1em,topsep=-.1em]
\item[2.1.1.]
For each unlabeled $1$-tree $G$, we look for the {couples} $f_i, f_j\in\F$ such that the current\x{ sequence} list $\G$ (let $|\G|=k$) still contains no graph that \separat{es} $f_i$ and $f_j,$ while $G$ \separat{es} them on some pair $\{u,v\}$ of its nodes.
\aB{We choose the most prior pairs, with
$\{u,v\}$ being \emph{prior\/} to $\{w,z\}$ iff
$(\delta(u,v)>\delta(w,z)$ or
 $\delta(u,v)=\delta(w,z)\wedge\tilde\delta(u,v)>\tilde\delta(w,z))$,
where
$\delta(u,v)=|D(u)-D(v)|$,
$\tilde\delta(u,v)=|\tilde D(u)-\tilde D(v)|$, and
$\tilde D(u)=\sum_{w:\:w\text{ is adjacent to }u} D(w)$.
}

\begin{enumerate}[itemsep=-.1em,topsep=-.0em]
\item[2.1.1.1.]
\x{In this case,}If $\{f_i,f_j\}$ is such a couple, then we add $G$ to the\x{ sequence} list $\G$ as $G_{k+1}$ and associate 
a separating triple $(G_{k+1},u,v)$ with\x{ the pair} $\{f_i,f_j\}$.
\end{enumerate}\end{enumerate}\end{enumerate}
\item[3.]
We stop the generation process when a separating triple $(G_{k+1},u,v)$ is associated 
with every pair of distinct centrality measures in~$\F$.
\end{enumerate}

\smallskip
\hlb{The number of nontrivial 1-trees of order $n$ is exponential, however, for $n=4\cdc10$ it is only $1,4,12,32,88,239,656,$ respectively} \cite{SloanUnicycles}. 
\hlb{All measures in the set $\Ff$ of $40$ centralities (see Section}~\ref{ss:TestSu}) \hlb{are distinguished by 27 separating triples taken from 12 1-trees of order $n\le7$ plus 2 triples from one 1-tree of order~8.}
Therefore, the ``separating power'' of $1$-trees on $8$ nodes is far from exhausted and can potentially separate a much larger number of measures.
Moreover, \x{The above procedure allows for variations. Say, }
for our test survey\x{the survey in Section~ \ref{ss:TestSu}} we limited ourselves to\x{ considered only} $1$-trees with node degrees\x{ less than five} at most four, which\x{ restriction} \x{does not apply}is not \x{required}\hla{obligatory} in the above procedure. \x{Moreover, in both procedures we excluded from consideration the trivial case of graphs with 2 or 3 nodes.}

However\x{On the other hand}, there is no guarantee that all\x{ possible} couples of\x{ different (rank-nonequivalent)} measures in $\F$ can be \separat{ed} by $1$-trees. 
To guarantee a complete separation, we propose the following general\x{universal:1/1000} procedure that iterates over the entire set of connected graphs.
\hlb{It systematically implements our idea of how to make answering survey questions as easy as possible. This idea boils down to the following.} 
\begin{enumerate}[itemsep=-.02em,topsep=-.1em]
\item[\paa{a}] \x{We believe that}\hlb{The most convenient graphs for experts are small 1-trees.}
\item[\paa{b}] \hlb{The second-best graphs are trees.}
\item[\paa{c}] \hlb{For other graphs, the lower the density, the better.} 
\item[\paa{d}] \hlb{For fixed-density graphs, the lower the order, the better.}
\item[\paa{e}] \hlb{Low density is preferred to low order.}
\end{enumerate}
This procedure includes\x{implements} the previous one\x{ above procedure in the first step} as the first stage to make the most of $1$-trees.

\begin{enumerate}[itemsep=-.08em,topsep=.0em]
\item[1.]
At the beginning, the list $\G$ is empty.
\item[2.]\label{i:2.2}
Increase $n$ starting with $3,$ and while there are unseparated measures in $\F$ and $n\le8,$ sequentially use all connected graphs with $n$ nodes and $n$ edges ($1$-trees) to separate measures in $\F$ as in the previous procedure.
\item[3.]
If there are still unseparated measures in $\F$, proceed as in Item~2\x{\ref{i:2.2}} with trees (i.e., with connected graphs having $n-1$ edges).
\item[4.]
If there are still unseparated measures in $\F$, proceed as in Item~2\x{\ref{i:2.2}} with connected graphs having $n+1$ edges.
\item[5.]
Continue in the following way until all\x{ the} measures in $\F$ are separated:
\begin{enumerate}[itemsep=-.10em,topsep=-.3em]
\item[5.1.]
For each increasing $n>8$\xy\x{ we}:
\begin{enumerate}[itemsep=-.1em,topsep=-.1em]
\item[5.1.1.]
Sequentially use, to separate the not yet separated measures in $\F,$ all connected graphs with $n$ nodes and: \paa{a}~$n$ edges ($1$-trees); \paa{b}~$n-1$ edges (trees); \paa{c}~$n+1$ to $2n-8$ edges;
\item[5.1.2.]\label{i:2last}
Generate and use: \paa{d}~all connected graphs with $n'\le n$ nodes and $n'+n-7$ edges, starting with the smallest $n'$ allowing that many edges and increasing~$n'.$
\end{enumerate}
\end{enumerate}
\end{enumerate}
%
%

\smallskip
In this general\x{universal} procedure, with each incremented\x{increasing} $n>8$, we increase\x{ by $1,$} the maximum difference between the number of edges and nodes to $(n'+n-7)-n'=n-7$ (see \paa{d} in 5\x{\ref{i:2last}}.1.2 above) and check all graphs with this\x{ maximum} difference and the number of nodes at most~$n.$ This maximum difference grows slowly to keep the test graphs simple enough by keeping their density as low as possible. 
%

The next task is constructing a questionnaire\x{ (in the form of a decision tree)} that allows the expert to choose the centrality measure that suits them best.


\subsection{Constructing a Decision Tree}
\label{ss:DecTree}

\noindent \x{In this section, w}We now present\x{ an algorithm} a procedure for constructing a questionnaire, which takes the form of a decision tree.

In every next\x{each} question, the expert is asked to compare the centrality of two nodes $u$ and $v$\x{: $f(u)$ and $f(v)$} in some graph~$G_k\in\G.$ This graph separates\x{ (on some pairs $\{u,v\}$ of its nodes)} certain measures that\x{ satisfy all the conditions} match all the expert's responses already received \hlb{(let the set of such measures be~$\F_x$)}. 
\x{Another}The new response is $f(u)>f(v),$ or $f(u)<f(v),$ or $f(u)=f(v)$ and it narrows down \hlb{$\F_x$}\x{the set of suitable measures}\x{ by preserving} keeping only those \hlb{$f_i\in\F_x$} for which this response is correct\x{ true}.
\hlb{The survey only offers answer options corresponding to some measures in~$\F_x$.}
The survey\x{ must} continues until only one measure remains. 

Thus the questions in the survey\x{ generally} depend on the\x{ answers} responses to the previous ones. Therefore, the questionnaire has the structure of a rooted directed tree. 
The expert navigates the tree answering the questions associated with the root and intermediate vertices.
Each answer is identified with an arc directed from the vertex corresponding to the question.
The \emph{leaves\/}, i.e., the vertices of the\x{ decision} tree that have no outgoing arcs, are identified with the\x{ resulting} measures. 

We now present a procedure\x{an algorithm} for constructing such a\x{ this directed} tree.
\x{We}Let us have a set $\F$ ($|\F|>1$) of centrality measures and a\x{ sequence} list of distinguishing graphs $\G=(G_1\cdc G_K).$ For each couple of distinct measures $f_i,f_j\in\F,$ a triple $(G_k,u,v)$ is associated with this couple such that $G_k\in\G$ \separat{es} $f_i$ and $f_j$ on the pair $\{u,v\}$ of its nodes.

At the {\em initial\/} step of the\x{ algorithm} procedure, we are at the root of the tree. There are no other vertices in the tree yet, and no question is associated with the root.

On {\em each\/} step of the\x{ algorithm} procedure, we are at some vertex $x$ of the tree.
A \emph{standard step\/} of the\x{ algorithm} procedure is any step, except for ``finish.''

The standard step consists of the following\x{ actions}.

\begin{itemize}[itemsep=-.08em,topsep=.11em]
\item
  Consider the path from the root to~$x.$ Each arc of this path corresponds to some condition: it is a response\x{n answer} to the question associated with the vertex this arc is directed from. Let $\F_x\subseteq\F$ be the set of measures satisfying all conditions in this path (if $x$ is the root, then $\F_x=\F$).
\item
  If no question has yet been associated with $x$ and ${|\F_x|>1}$, then take $G_k\in\G$ with the smallest $k$ such that $G_k$ \separat{es} some measures in $\F_x$ on a certain pair of its nodes $\{u,v\}.$
  \aB{As well as in Section~\ref{ss:Genera}, we choose the most prior pair $\{u,v\}$ of this kind} and associate the question ``What inequality (equality) holds true for $f(u)$ and $f(v)$ in $G_k$?'' with~$x.$ Draw two or three arcs directed from $x$ to newly created vertices, depending on which of the three conditions $f(u)>f(v),$ $f(u)<f(v),$ or $f(u)=f(v)$ are met by the measures in~$\F_x.$ Assign the conditions that can be satisfied by the measures in~$\F_x$ to these arcs. Mark these arcs as ``new.'' 
\item
  If there is at least one ``new'' arc directed from $x,$ choose any such arc, move to the vertex this arc is directed to, and mark this arc as ``old.''
\item
  If $|\F_x|=1,$ then associate with $x$ the unique centrality measure that belongs to~$\F_x.$
\item
  If \paa{a} there is no ``new'' arc directed from $x,$ while there is a question associated with $x,$ and $x$ is not the tree root or \paa{b}~$|\F_x|=1,$ then make one move back from $x$ toward the root. 
\item
  If there is no ``new'' arc directed from $x,$ while there is a question associated with $x$ and $x$ is the root, then finish: the decision tree is built. Otherwise, make another standard step from~$x.$
\end{itemize}

A high-level pseudocode of the standard step is given\x{ shown} in Algorithm~\ref{a_Tree}.
\hlb{Its worst-case complexity is $O(|\F|^3)$.}

\begin{algorithm}[th]
  \caption{Recursive Construction of a Decision Tree.\label{a_Tree}}
  \begin{algorithmic}
  \State \textbf{Input:} $\mathcal{F}$, a set of centrality measures; $\mathcal{G}=(G_1,\dots,G_K)$, a list of distinguishing graphs; a set of distinguishing triples of the form $(G_k,u,v)$.
  \State \textbf{Result:} Decision tree for choosing a centrality measure
  \Procedure{standardStep}{$x$}
      \If{$|\mathcal{F}_x|>1$}
        \State $(G_k,u,v) \gets \text{distinguishMeasures}(\mathcal{F}_x)$
        \State $x.\text{associateQuestion}(\text{``$f(u)$ vs $f(v)$ in $G_k$''})$
        \State $x.\text{addNewArcsAndChilds}(G_k,u,v)$
        \While{$x.\text{hasNewArcs()}$}
          \State $arc \gets x.\text{getAnyNewArc()}$
          \State $\text{markArcAsOld}(arc)$
          \State $\text{standardStep}(arc.target)$
        \EndWhile
      \Else
        \State $x.\text{associateUniqueMeasure}(\mathcal{F}_x)$
      \EndIf
  \EndProcedure
  \end{algorithmic}
\end{algorithm}

\subsection{Extending a Decision Tree}
\label{ss:ExtDecTree}

\noindent Suppose now that several new centrality measures (for example,\x{ those that have} recently appeared in the literature)\x{ have been} are added to an existing set~$\F$.
Should we rebuild\x{redesign} the\x{ questionnaire} decision tree from the beginning? The answer is no.

\hla{First} consider the simplest case where only\x{just} one measure $f'$ is added. There are two possibilities:

\begin{itemize}[itemsep=-.1em,topsep=-.1em]
\item[\paa{a}]
  the new measure $f'$ satisfies the same set of conditions (answers\x{responses1/3} to the questions of the current decision tree\x{questionnaire}) 
  as one of the ``old'' measures;
\item[\paa{b}]
  in the current decision tree\x{ of the current questionnaire} applied to the extended set $\F'=\F\cup\{f'\},$ there is a vertex $x$ 
  such that $f'\in\F'_x,$ but only two arcs are directed from $x,$ while the third possible answer to the question associated with $x$ is correct\x{true} for~$f'.$
\end{itemize}

In case \paa{b},\x{ the only thing to do is} it remains\x{ only} to add a\x{the} third arc directed from $x,$ i.e.,\x{ to} attach the third possible answer to the question associated with $x,$ after which an\x{the} updated\x{ decision} tree is \hla{built.}

In \hla{case} \paa{a}, we need to distinguish $f'$ from an\x{ the} ``old'' measure $f_i$ that satisfies the same set of conditions. Since $f_i$ and $f'$ end up on the same leaf of the current decision tree, we need to make this leaf an intermediate vertex by associating a new question with it. To formulate this question, we first check if $\G$ contains a graph that \separat{es} $f_i$ and $f'$ on some pair of its nodes $\{u,v\}.$ If such a graph $G$ exists, then the required question is: ``Which inequality (or equality) is true for $f(u)$ and $f(v)$ in~$G$?'' Otherwise, we need\x{first have to generate} a new graph that \separat{es} $f_i$ and $f',$ so we call the graph generation procedure described in \x{the {\em Generating Graphs that Distinguish Measures\/} \hla{section}}Section~\ref{ss:Genera} for the set of measures $\{f_i,f'\}$.

\aBa{Thus, adding a new measure $f'$ leads to the addition of\x{adding} either a question or a possible answer to an existing question. 
\x{This proves, among others, that}So the total number\x{ $q$} of questions in the questionnaire does not exceed $m-1,$ where $m$ is the number of measures. If three arcs are directed from $k$ vertices of the tree, then the number of questions is obviously $m-k-1.$} 

\hlb{The average survey length for an expert can obviously vary from $\log_3m$ (when $m=3^b$, $b\in\N$, and each question splits the remaining measures into $3$ groups of equal size) to $\frac{(m+2)(m-1)}{2m}$ (when each question isolates just one measure).} 

\hlb{This implies that the number $s$ of separating triples for $m$ measures satisfies $\log_3m\le s\le m-k-1$.}

Suppose now that {\em several\/} measures have been added to~$\F.$
Then we need the following:
\begin{enumerate}[itemsep=-.05em,topsep=.08em]
\item[1.]
  For each new measure $f',$ check the above condition \paa{b}; if it is satisfied, then add the lacking arc to the decision tree\x{ as for the case~\paa{b}}; 
\item[2.]
  Attach the new measures for which condition \paa{b} is violated to the existing leaves of the tree,
  as for the above case~\paa{a};
\item[3.]
  Check whether $\G$ \hla{contains} graphs that \separat{e} the measures associated with the same leaf and add the separating questions whenever such graphs exist;
\item[4.]
  For the couples of measures that cannot be distinguished by \hla{any}\x{the} graphs in $\G,$ generate new graphs as described in \x{the {\em Generating Graphs that Distinguish Measures\/} section}Section~\ref{ss:Genera} and add the required questions.
\end{enumerate}

The above procedures\x{ described in Sections~ \ref{ss:Genera} to \ref{ss:ExtDecTree} above implement form the algorithmic basis of} of the culling method allow one to \x{generate}build a decision tree\x{survey designed to select} for selecting the most appropriate centralities\x{ measures} for\x{from} any set of measures~$\F.$
\hlb{They generate a survey that starts with questions on fairly small, low-density graphs and gradually increases their order and size. As seen in Section}~\ref{ss:TestSu}, \hlb{the first questions separate the ``semantic'' classes of measures, and the following questions isolate the subclasses within them.} 

\subsection{Test Survey} 
\label{ss:TestSu}

\noindent In {\em Materials and Methods}\x{Section~\ref{s:MatMet}}, we present\x{consider (introduce)} several parametric classes of centralities based on similarity/dissimi\-larity measures for network nodes.
Representatives of these classes are included in the set $\Ff$ designed\x{used} to test the culling method. This set consists of the\x{ following forty} centralities listed in Table~\ref{t:ListMeasures}.
\x{Before presenting a test survey, we introduce several parametric classes of\x{new} centralities.
They are based on similarity/dissimilarity measures for network nodes; representatives of these classes will be included in the survey.
To test the culling\x{proposed approach} method, consider the set $\Ff$ consisting of the\x{following} centralities listed in Table~\ref{t:ListMeasures}.
}

\begin{table}[t!] 
\centering
\caption{The test set~$\Ff$ of centrality measures\label{t:ListMeasures}}

\smallskip
\begin{tabular}{ll}
\jJ{~\,1.~Betweenness\!\cite{freeman1977set}}&               \jJ{21.~Bonacich\!\cite{Bonacich87}}\\
\jJ{~\,2.~Closeness\!\cite{bavelas1948\x{,bavelas1950}}}&    \jJ{22.~Total communicability\!\cite{benzi2013total}}\\
\jJ{~\,3.~Connectivity\!\cite{Khmelnitskaya23}}&            \jJ{23.~Communicability $(K_{ij})$\!\cite{estrada2005subgraph}}\\
\jJ{~\,4.~Connectedness\,power\!\cite{Mago18Power}} \!\!\!\!\!\!\!\!\!\!&\jJ{24.~Walk $(K_{ij})$\!\cite{Katz53}}\\
\jJ{~\,5.~Degree\!\cite{bavelas1948\x{,proctor1951analysis}}}& \jJ{25.~Walk $(K_{ii})$\!\cite{Katz53}}\\
\jJ{~\,6.~Coreness\!\cite{bae2014identifying}}&            \jJ{26.~Estrada\!\cite{estrada2005subgraph}}\\
\jJ{~\,7.~Bridging\!\cite{breitling2004rank}}&             \jJ{27.~Eigencentrality\,(Dice dissimilarity\!\cite{alvarez2015eigencentrality})}\\
\jJ{~\,8.~PageRank\!\cite{BrinPage1998anatomy}}&           \jJ{28.~Eigencentrality\,(Jaccard\,dissimilar.\!\cite{alvarez2015eigencentrality})}\\
\jJ{~\,9.~Harmonic\,closeness\!\cite{marchiori2000harmony\x{,boldi2014axioms}}}&\jJ{29.~Closeness\,(Forest\!\cite{CheSha01})}\\
\jJ{10.~Eccentricity\!\cite{bavelas1948\x{,harary1953}}}&    \jJ{30.~Closeness\,(Heat\!\cite{KondorLafferty02diffusion})}\\
\jJ{11.~$p$-Means\,$(p=-2)$\!\cite{deAndrade2019p}}&      \jJ{31.~Closeness\,(logarithmic Forest\!\cite{Che11DAM})}\\
\jJ{12.~$p$-Means\,$(p=0)$\!\cite{deAndrade2019p}}&       \jJ{32.~Closeness\,(logarithmic Walk\!\cite{Che12DAM})}\\
\jJ{13.~$p$-Means\,$(p=2)$\!\cite{deAndrade2019p}}&       \jJ{33.~Closeness\,(logarithmic Heat\!\cite{KondorLafferty02diffusion\x{,Che13Paris}})}\\
\jJ{14.~Beta current flow\!\cite{avrachenkov2015beta}}&    \jJ{34.~Closeness\,(log-Communicability\!\cite{ivashkin2016logarithmic})}\\ 
\jJ{15.~Weighted degree\!\cite{bandyopadhyay2017generic-}}& \jJ{35.~Eccentricity\,(Forest\!\cite{CheSha01})}\\
\jJ{16.~Decaying degree\!\cite{bandyopadhyay2017generic-}}& \jJ{36.~Eccentricity\,(Heat\!\cite{KondorLafferty02diffusion})}\\
\jJ{17.~Decay\!\cite{tsakas2018decay}}& \jJ{37.~Eccentricity\,(logarithmic Forest\!\cite{Che11DAM})}\\
\jJ{18.~Generalized degree\!\cite{csato2017measuring}}&    \jJ{38.~Eccentricity\,(logarithmic Walk\!\cite{Che12DAM})}\\
%
\jJ{19.~Katz\!\cite{Katz53}}&                              \jJ{39.~Eccentricity\,(logarithmic Heat\!\cite{KondorLafferty02diffusion\x{,Che13Paris}})}\\
\jJ{20.~Eigenvector\!\cite{landau1895relativen\x{,bonacich1972factoring}}}&\jJ{40.~Eccentricity\,(log-Communicability\!\cite{ivashkin2016logarithmic})} 
\end{tabular}
\end{table}

The set $\Ff$ contains several\x{many} pairs of closely related centralities. \x{The main purpose of}We include them to check whether they can be separated by sufficiently small low-density graphs, or whether large or dense graphs are necessary for this purpose.

The list $\G$ of distinguishing graphs generated\x{constructed (obtained) (provided) by the culling method} for $\Ff$ consists of $13$ $1$-trees shown in Fig.~\ref{f:SepGraphs}.
\hlb{The culling decision tree involves $29$ separating triples.}\x{ questionnaire consists of $33$ questions in total.} The length of a particular survey ranges from 1 to 12, with an average\x{ length} of $6.6$.\x{ per measure.}

\begin{figure}[!t]
\begin{center}
\includegraphics[width=.7\linewidth]{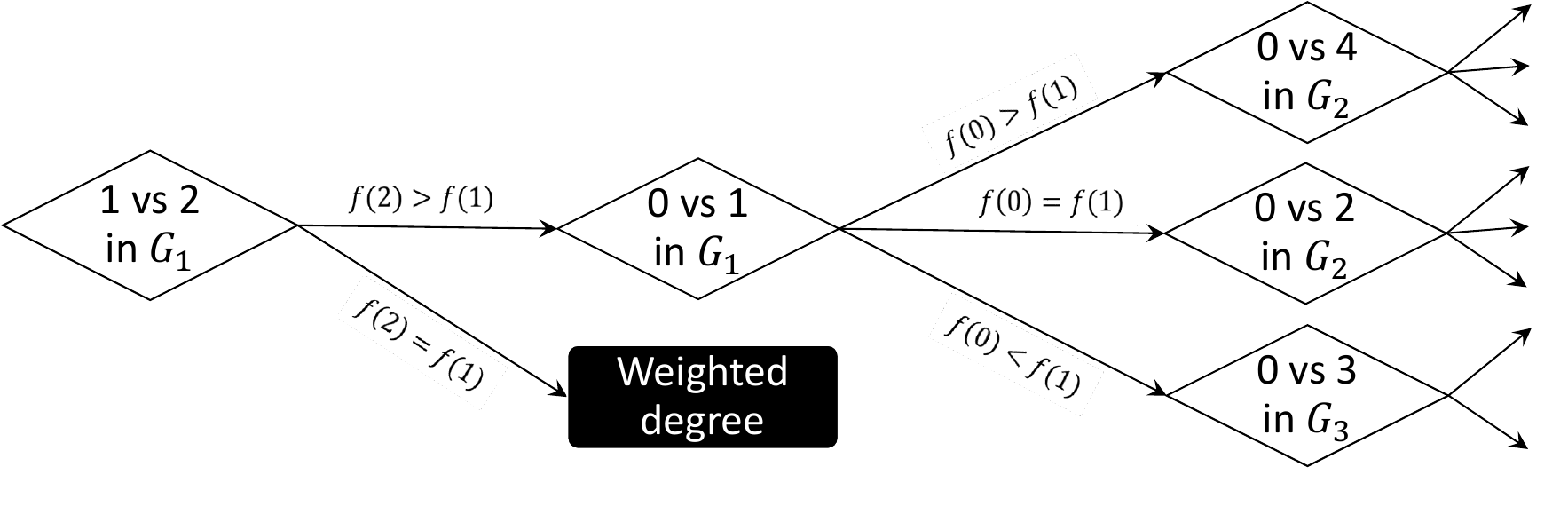}\\ 
\smallskip
\includegraphics[width=.6\linewidth]{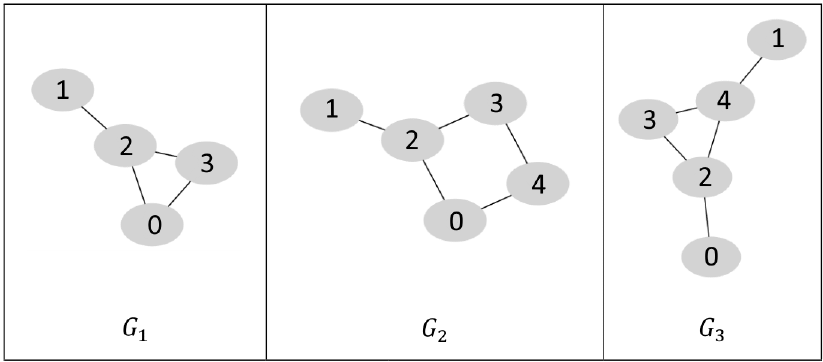} 
\caption{The beginning of the decision tree for the test set $\Ff.$\x{ of 40 measures.} \label{f:TreeBegin}}
\end{center}
\end{figure}

The survey begins with 
the question about the centrality of nodes 1 and 2 in $G_1$ (Fig.~\ref{f:TreeBegin}).
If the expert chooses\x{ answers,says responds that option} $f(2)>f(1),$ then the second question concerns\x{ is on} the centrality of nodes 0 and 1 in the same graph. If the answer\x{ to it (this question)}
            is $f(0)>f(1),$ then the third
    question is about the centrality of nodes 0 and 4 in $G_2$. \x{Alternatively}Otherwise, \hla{the expert who chooses} $f(0)=f(1)$
    is asked to compare $f(0)$ and $f(2)$ in $G_2,$ whereas in the case of answer $f(0)<f(1),$ the third
    question is about the centrality of nodes 0 and 3 in~$G_3$. If the expert chooses\x{ answers the first question that}
               $f(2)=f(1)$ for the first question, then the survey ends\x{ with the result} yielding the Weighted degree\x{c}~\cite{bandyopadhyay2017generic-} centrality\x{ measure}.\x{: it is the only index\x{ one (measure)} in $\Ff$ that meets this condition.}

\begin{figure*}[!t]
\begin{center}
\includegraphics[width=9.8cm]{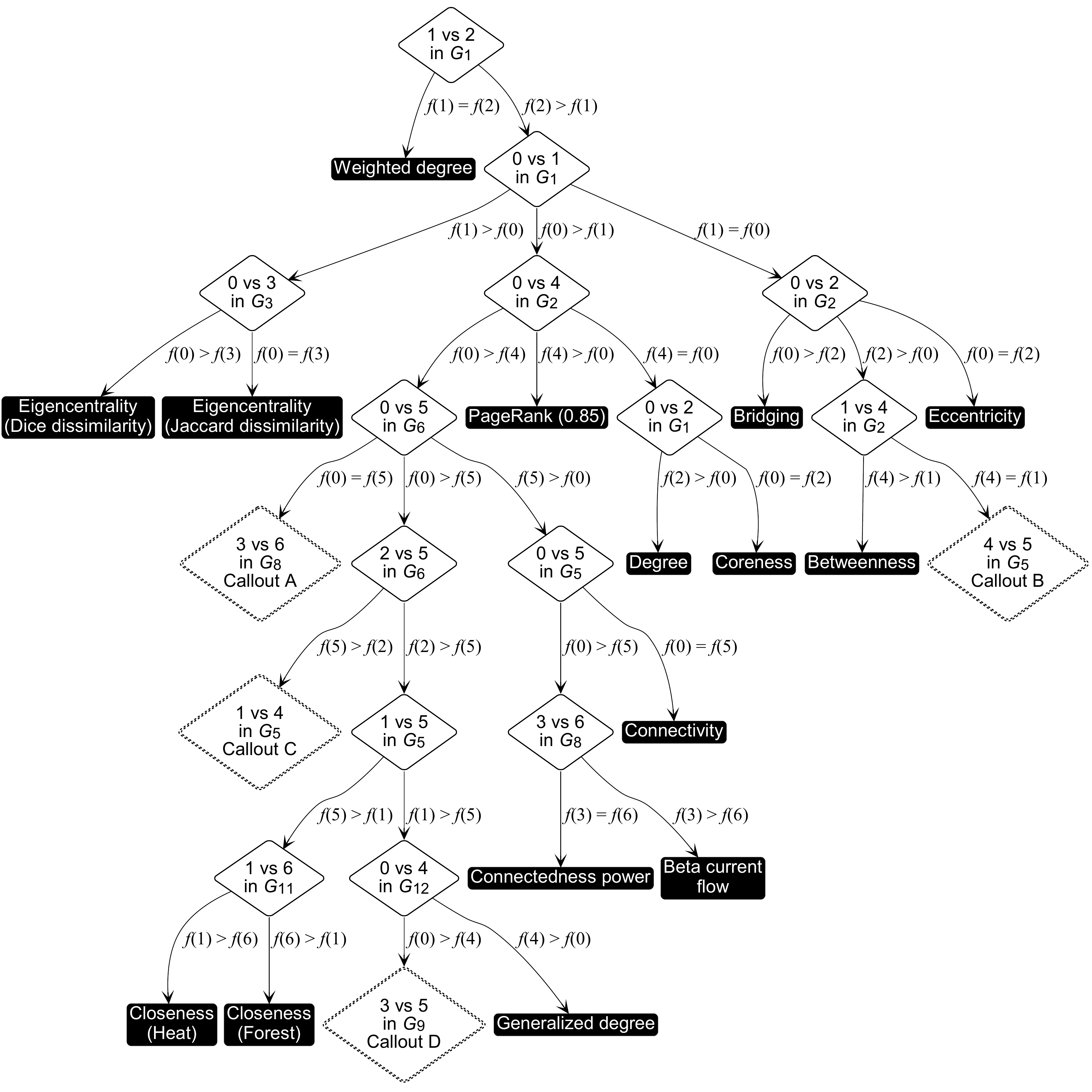} 
\caption{The decision tree for the test set $\Ff$; the callouts A to D are shown in Fig.~\ref{f:STree}.\x{ of 40 measures.} \label{f:Tree}}
\end{center}
\end{figure*}

\begin{figure*}[!t]
\begin{center}
{\scriptsize
\includegraphics[height=5.77cm]{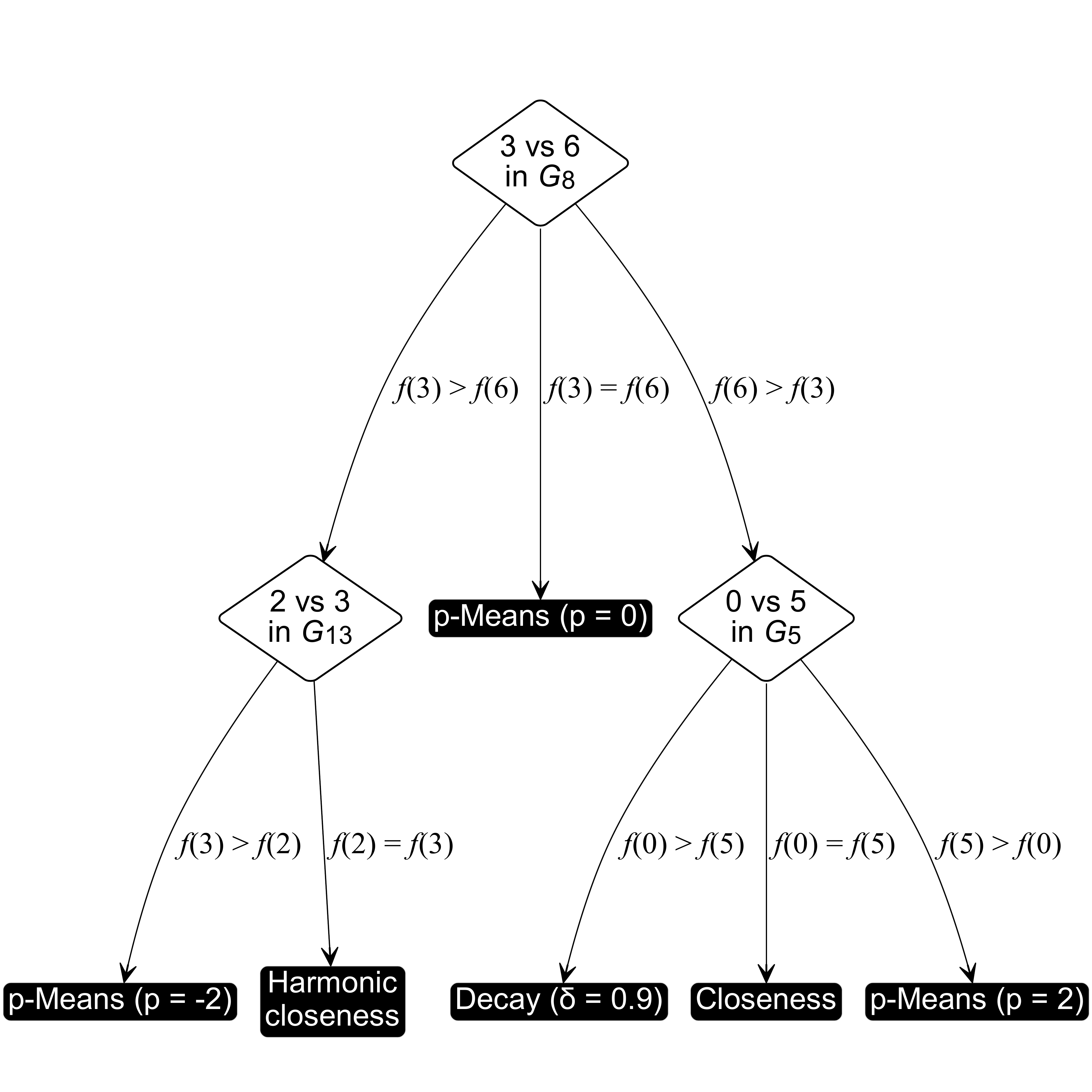}$\quad$  
\includegraphics[height=5.47cm]{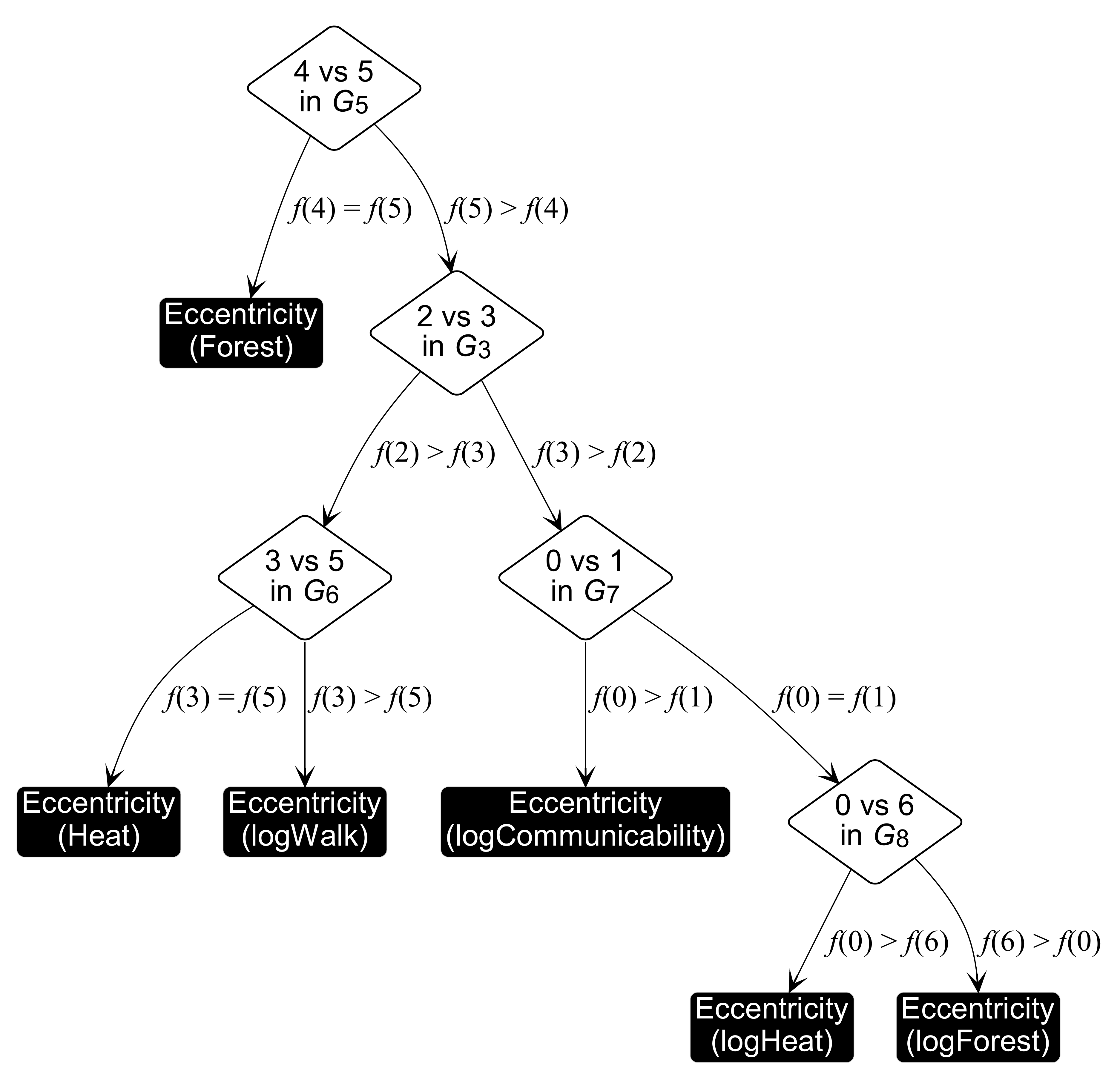}         

Callout A\hspace{16.58em}Callout B            
\bigskip

\includegraphics[height=5.70cm]{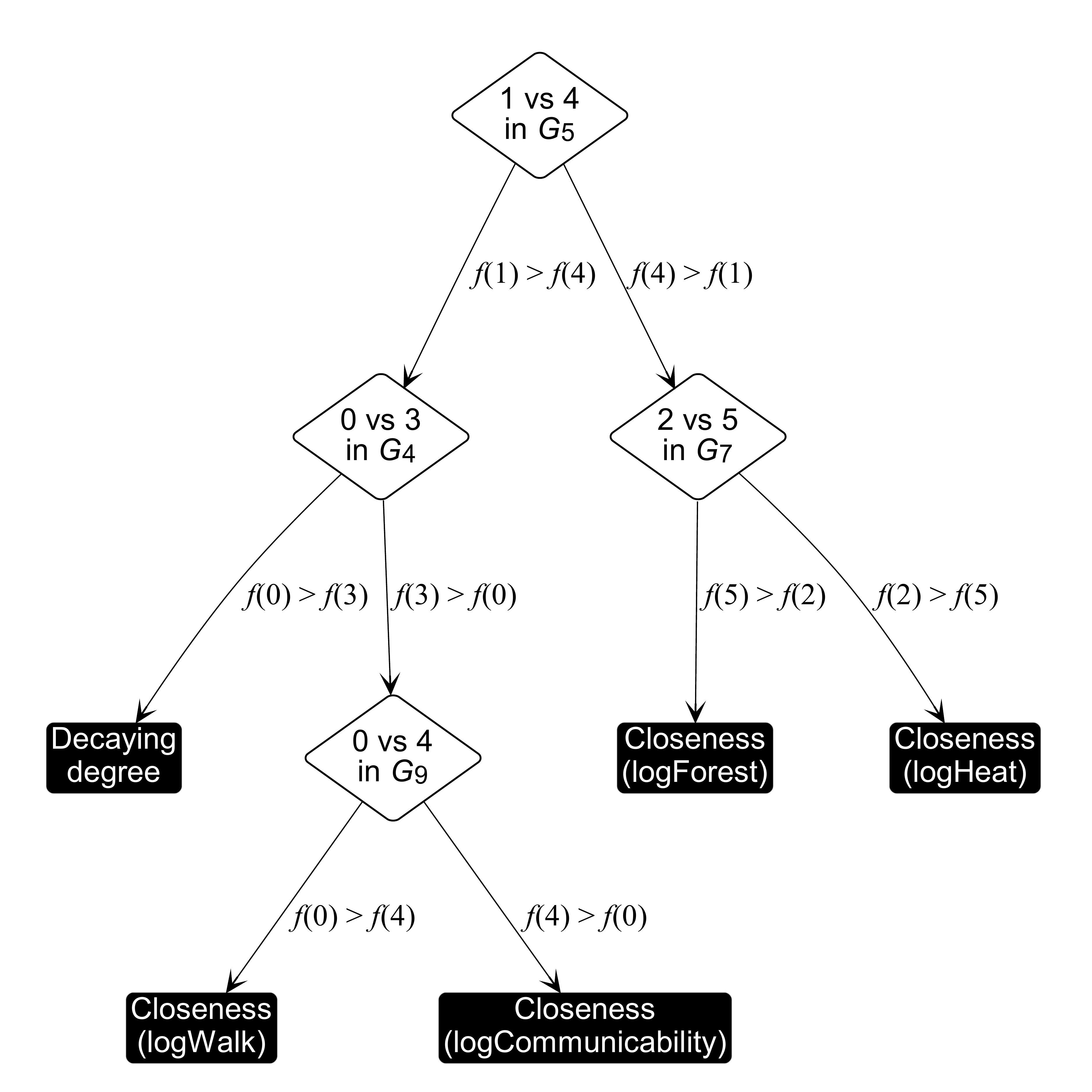}$\quad$  
\includegraphics[height=5.70cm]{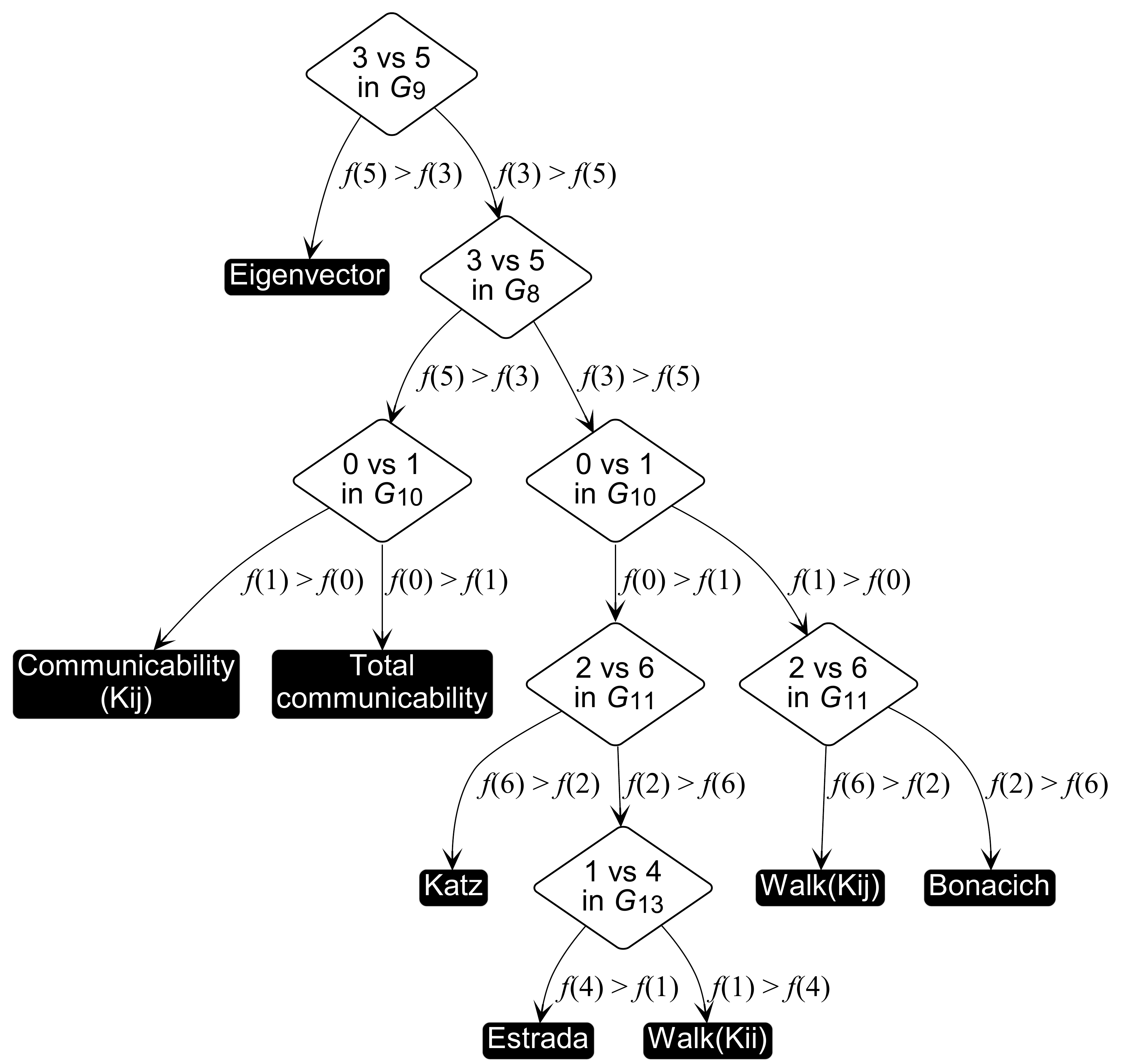}         

Callout C\hspace{16.58em}Callout D}           
\caption{Callouts A to D in Fig.~\ref{f:Tree}. 
\label{f:STree}}
\end{center}
\end{figure*}

\x{Thus, }The beginning of the survey\x{ decision tree} is represented by the\x{ rooted} tree of Fig.~\ref{f:TreeBegin}. The whole tree is \x{presented}shown in Fig.~\ref{f:Tree} and Fig.~\ref{f:STree}.

\begin{figure}[!t]
\begin{center}
\includegraphics[width=.6\linewidth]{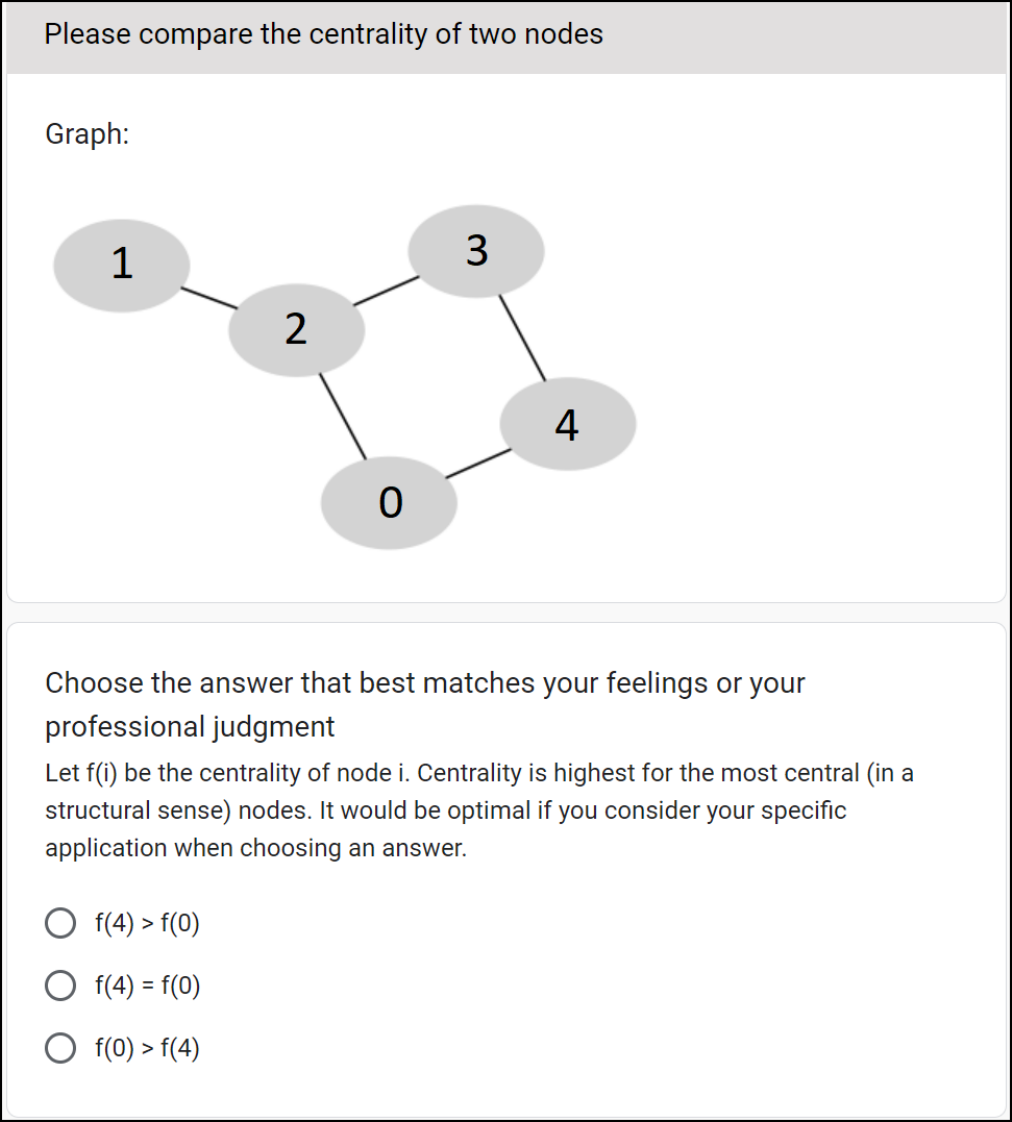} 
\caption{A sample survey form in \cite{OurWebSurv19}.\label{f:form}}
\end{center}
\end{figure}

\x{A survey based on this decision tree} The corresponding survey is\x{was} implemented using Google Forms~\cite{OurWebSurv19}.
It allows everyone to choose from $\Ff$ the measure that best suits\x{matches} their concept of centrality\x{beliefs, preferences}.
Each survey form\x{question} (\x{a sample}one of which is shown in Fig.~\ref{f:form}) contains\x{is accompanied by} the instruction: ``Choose the answer that best matches your feelings or your professional judgment... It would be optimal if you consider your specific application when choosing an answer.''

Of particular interest will be the statistics of measure selection by a significant number of experts.
The leaders of the early limited statistics are\x{were} Degree, Closeness (Heat), and Closeness (logarithmic Forest).

Note that a decision tree for the entire set of known centralities can be built by extending the tree (or, in other terminology, the {\em dendrogram}) presented in Figures~\ref{f:Tree} and~\ref{f:STree}.

\subsection{Combination with the Axiomatic Approach} 
\label{ss:Combining}

\noindent It can be seen\x{observed} that some\x{ of the(1/2)} questions in the above test survey are quite simple. 
Say, \x{the1/2}comparison of the centrality of nodes 1 and 2 in $G_1$ (Fig.~\ref{f:TreeBegin}) should be clearly in favor of~2. Therefore, the first question along with the\x{ measure} Weighted degree\x{c} index assigning equal centrality to nodes 1 and 2 can be eliminated, which reduces the number of questions in each\x{ particular} survey by one.

The second question on\x{about} the comparison of $f(0)$ and $f(1)$ in $G_1$ is less obvious, but also fairly simple. Indeed, the answer $f(0)<f(1)$ is unlikely to be popular, and most experts would probably choose\x{prefer} $f(0)>f(1).$ Based on this, the Eigencentrality measures based on Dice dissimilarity\x{c} and Jaccard dissimilarity\x{c} can be eliminated as they \x{set}suggest $f(0)<f(1).$ The answer $f(0)=f(1)$ \x{is unlikely to find many supporters, however,} \hla{may have some basis} and \x{ the set} $\Ff$\x{ of measures} contains nine centralities that match\x{support} it. These are the well-known Bridging\x{c} and Betweenness\x{c} measures and also the Eccentricity\x{c} measures based on the Shortest path, Forest, Heat, logarithmic Forest, logarithmic Heat, logarithmic Walk, and logarithmic Communicability proximities. The decision tree of Fig.~\ref{f:Tree} makes the expert realize that choosing either\x{any} of them\x{se measures} implies adopting that nodes 0 and 1 in $G_1$ are equally central\x{have equal centrality values}. Furthermore, the subsequent question (comparing nodes 0 and 2 in $G_2$) reveals that Bridging\x{c} states that $f(0)>f(2),$ which is not very easy to accept; $f(0)=f(2)$ leads to Eccentricity\x{c}; the remaining seven measures on this branch of the tree suggest $f(0)<f(2).$ 

Presumably\x{ Supposedly} the most popular answer, $f(0)>f(1),$ to the second question is followed by\x{leads to} the comparison of $f(0)$ and $f(4)$ in~$G_2.$ The rather specific answer $f(0)<f(4)$ leads to the PageRank\x{c} measure
(its peculiarities, which occur\x{hold} for all $\alpha\in\,(0;1)$, were discussed in \x{ Introduction}\x{ the introductory section}Section~\ref{s:intro} and in more detail in~\cite{Che23JCNCentr}),
$f(0)=f(4)$ to Degree\x{c} \hla{and}\x{or} Coreness\x{c}, while $f(0)>f(4)$ to the remaining 25 measures on this branch of the tree. 

The presence of\x{ fairly simple} questions\x{ in the survey}, the answer to which is prompted by common sense or an application-specific  understanding, suggests the formulation of\x{formulating} the corresponding properties of centrality measures in the form of normative conditions (axioms). 
If the expert\x{ user} deems some of them necessary, this can \x{dramatically}drastically reduce the set of measures to be compared (cf.~\cite{Boldi23Monotonicity}). Based on this, we can offer the expert\x{user} an abbreviated\x{a special (personalized personal individual)} survey on the corresponding reduced set~$\F'\subset\F.$ 

This leads to a combination of culling with the axiomatic\x{ a normative} approach.
In the next\x{following} section, we discuss two\x{a couple of} conditions that can be used in\x{ such a} this combination.

\subsection{Abbreviated Surveys} 
\label{ss:Abbrev}

\noindent Among the axioms used in the literature to characterize centrality measures, the ones that are particularly appealing are those based on ordinal conditions. These axioms allow us to compare the centrality of certain nodes, but they do not provide explicit guidelines on how to calculate centrality in the general case.

\x{In this section, we}Let us consider two such axioms and show how they can help significantly reduce the culling survey.

First, one may believe that
the vector of centrality values of the \emph{neighbors\/} (adjacent nodes) of any\x{ arbitrary} node $u$ carries a lot of information about the centrality of $u$ itself
(cf. Consistency in~\cite{dequiedt2017local}).
A refinement of this requirement is that \emph{the greater the centrality values of the neighbors of a node$,$ 
the greater the centrality of the node itself}.


The following axiom is based on this idea. In the case of directed graphs, it appeared in \cite{CheSha97a\x{,CheSha99}}; for undirected graphs, in \cite{bandyopadhyay2017generic-,bandyopadhyay2018generic} under the name of Structural consistency. 

Let $N_u$ denote the set of neighbors of node $u$ in~$G.$

\axiom{Self-consistent monotonicity}{\!\!If for $u,v\in V,$ there is an injection from $N_u$ to $N_v$ such that every element
of $N_u$ is, according to $f(\cdot),$ no more central than the corresponding element of $N_v$, then $f(u)\le f(v).$ 
If$,$ additionally$,$ $|N_u|<|N_v|$ or ``no more'' is actually ``less'' at least once$,$ then $f(u)<f(v).$ 
}

For the subsequent analysis, we \hla{choose}\x{select} the following weaker and simpler axiom (cf.~\cite{CheSha98AOR}).

\axiom{Self-consistency}{\!\!If for $u,v\in V,$ there exists a bijection between $N_u$ and $N_v$ such that each element of $N_u$ is, according to $f(\cdot),$ no more central than the corresponding element of $N_v$, then $f(u)\le f(v).$ 
If ``no more'' is actually ``less'' at least once, then $f(u)<f(v).$}

The idea of\x{ the second} another axiom~\cite{skibski2018axioms} is quite different. It allows\x{ us} to compare the centrality of two endpoints of a bridge\x{ in~$G$}. 
Recall that we consider central\-ities \x{y measures}defined for all connected graphs $G$ with $|V(G)|>3.$  

%
\axiom{Bridge axiom}{\!\!If\x{ $G$ is connected and} the removal of edge $\{u,v\}$ from $E(G)$ separates\x{ splits}
$G$ into two connected components with node sets \hla{denoted by} $V_u\ni u$ and $V_v\ni v$ (i.e., $\{u,v\}$ is a {\em bridge}\x{ in $G$}), then $|V_u|<|V_v|\ToTo$ $f(u)<f(v).$}

A strengthening of this axiom is the Ratio property~\cite{Khmelnitskaya23}, which holds when for a positive $f(\cdot),$ under the same premise, $f(u)/f(v)=|V_u|/|V_v|.$

\hlb{For more about these axiom, including the rationale behind choosing them, a necessary and sufficient condition for Self-consistency, and a sufficient condition for the Bridge axiom, see} \cite{Che23JCNCentr}. \hlb{In particular,} it is shown that in the presence of Monotonicity and Equivalence, these axioms are incompatible.

\begin{figure}[!t]
\centering
\includegraphics[width=0.42\linewidth]{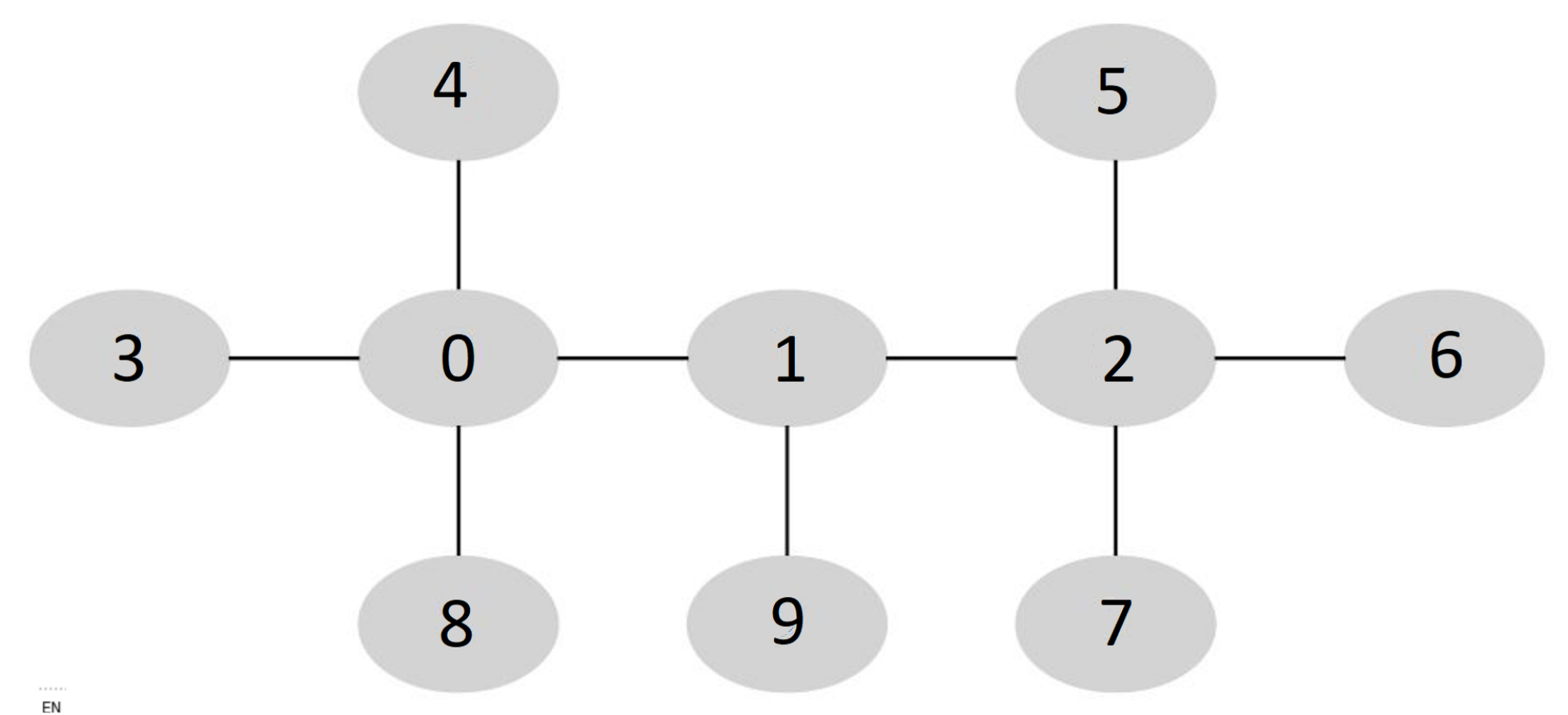}\\
{\footnotesize (a)}

\bigskip
\includegraphics[width=0.7\linewidth]{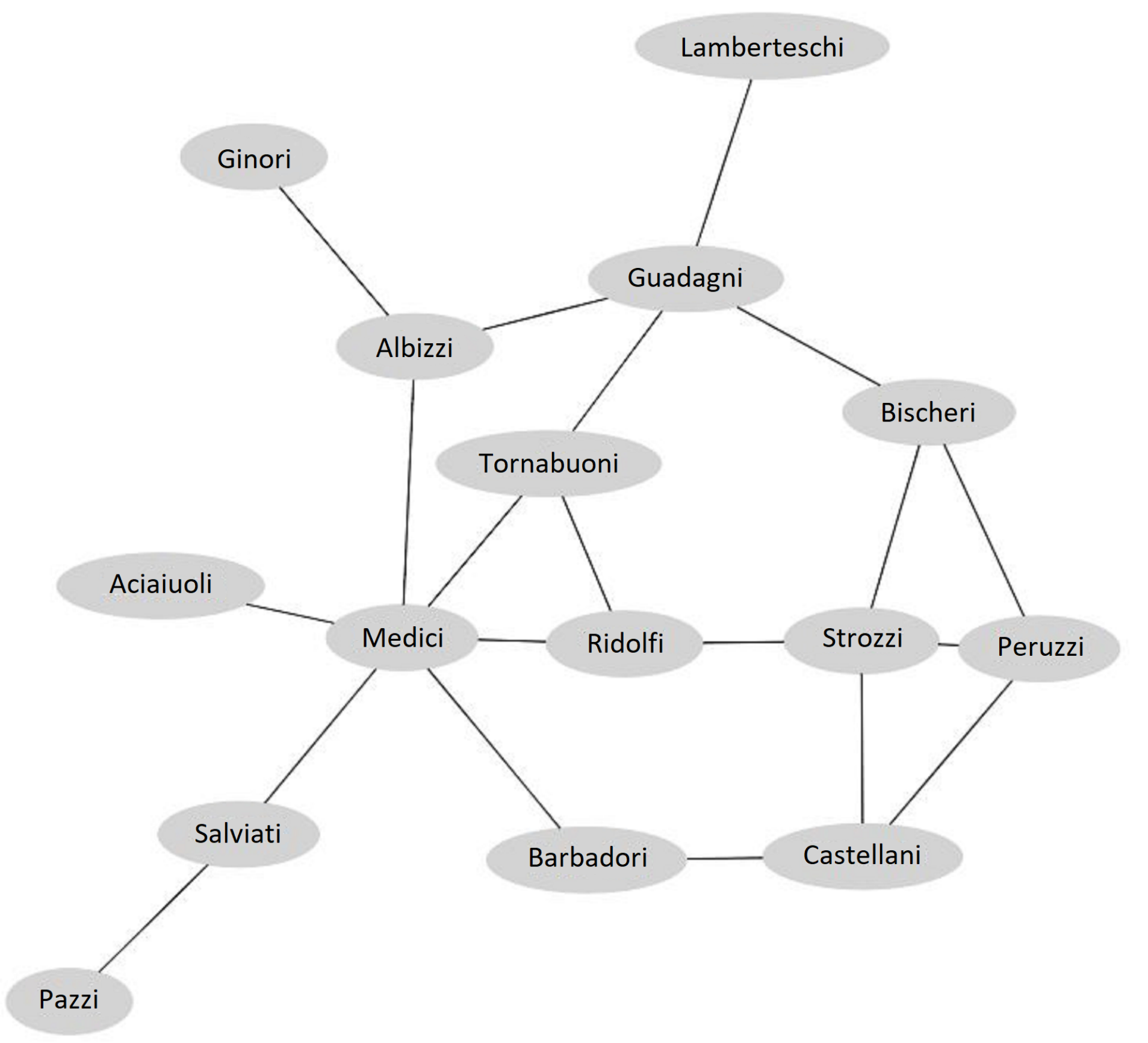}\\ 
\Up{.6}
{\footnotesize (b)}
\caption{Graphs used to test the Self-consistency and Bridge axioms.\\ (a)~Caterpillar~$G_{14}$. 
(b)~The Renaissance Florentine families marriage network $G_{15}$~\cite{PadgettAnsell93}.\label{f:Florenc}}
\end{figure}

\begin{table}[!ht]
\centering{
\caption{Verification of the Self-consistency and Bridge axioms for the measures in $\Ff$. `$\bm+$' indicates that the measure satisfies the axiom; `$\bm{+/-}$' that it is satisfied by a slightly modified measure; `$\bm-$' that the axiom is violated. In the latter case, a graph and an ordered pair of nodes are indicated for which the axiom is violated. ``Comm.'' is short for Communicability. \label{ta_C-A}}\x{\Up{1}} 

\smallskip
\begin{tabular}{lcc} 
\hline
{\jJ{Centrality measures}}                                &\jJ{Self-consistency}& \jJ{Bridge axiom}\\
\hline
\jI{~\:1.~Betweenness                                                    }&$-_{(0,3)G_4}  $&$-_{(0,5)G'_5\x{(G_5\ssm\{\xz\{1,2\}\xz\})}}$\\[-1.4pt]
\jI{~\:2.~Closeness                                                      }&$-_{(0,3)G_4}  $&$      +     $\\[-1.4pt]
\jI{~\:3.~Connectivity                                                   }&$-_{(0,3)G_4}  $&$      +     $\\[-1.4pt]
\jI{~\:4.~Connectedness power                                            }&$-_{(0,3)G_4}  $&$-_{(0,5)G_5}$\\[-1.4pt]
\jI{~\:5.~Degree                                                         }&$-_{(0,4)G_2}  $&$-_{(0,5)G_5}$\\[-1.4pt]
\jI{~\:6.~Coreness                                                       }&$-_{(0,1)G_7}  $&$-_{(0,5)G_5}$\\[-1.4pt] 
\jI{~\:7.~Bridging                                                       }&$-_{(0,3)G_4}  $&$-_{(0,5)G_5}$\\[-1.4pt]
\jI{~\:8.~PageRank\x{,\, $\alpha=0.85$}                                  }&$-_{(0,3)G_4}  $&$-_{(0,5)G_5}$\\[-1.4pt]
\jI{~\:9.~Harmonic closeness                                             }&$-_{(0,3)G_4}  $&$-_{(0,5)G_5}$\\[-1.4pt]
\jI{10.~Eccentricity                                                     }&$-_{(0,4)G_2}  $&$-_{(0,5)G_5}$\\[-1.4pt] 
\jI{11.~$p$-Means,\, $p=-2$                                              }&$-_{(0,3)G_4}  $&$-_{(0,5)G_5}$\\[-1.4pt]
\jI{12.~$p$-Means,\, $p=0$                                               }&$-_{(0,3)G_4}  $&$-_{(0,5)G_5}$\\[-1.4pt]
\jI{13.~$p$-Means,\, $p=2$                                               }&$-_{(0,3)G_4}  $&$-_{(0,5)G_5}$\\[-1.4pt]
\jI{14.~Beta current flow                                                }&$-_{(0,3)G_4}  $&$-_{(0,5)G_5}$\\[-1.4pt]
\jI{15.~Weighted degree                                                  }&$-_{(0,4)G_2}  $&$-_{(1,2)G_1}$\\[-1.4pt]
\jI{16.~Decaying degree                                                  }&$-_{(1,5)G_5}  $&$-_{(0,5)G_5}$\\[-1.4pt]
\jI{17.~Decay\x{,\, $\delta=0.9$}                                        }&$-_{(0,3)G_4}  $&$-_{(0,5)G_5}$\\[-1.4pt]
\jI{18.~Generalized degree\x{,\, $\ve=2$}                                }&$      +       $&$-_{(0,5)G_5}$\\[-1.4pt]
\jI{19.~Katz\x{,\, $a=(\rho(A)+1)^{-1}$}                                 }&$      +       $&$-_{(0,5)G_5}$\\[-1.4pt]
\jI{20.~Eigenvector                                                      }&$      +       $&$-_{(0,5)G_5}$\\[-1.4pt]
\jI{21.~Bonacich\x{,\, $\alpha=1,\,\beta=(\rho(A)+0.5)^{-1}$}            }&$      +       $&$-_{(0,5)G_5}$\\[-1.4pt]
\jI{22.~Total communicability\x{,\, $t=1$}                               }&$-_{(8,9)G_{14}}$&$-_{(0,5)G_5}$\\[-1.4pt]
\jI{23.~Communicability$(K_{ij})\x{,\, t=1}$                             }&$-_{(8,9)G_{14}}$&$-_{(0,5)G_5}$\\[-1.4pt]
\jI{24.~Walk$(K_{ij})\x{,\, t=(\rho(A)+1)^{-1}}$                         }&$-_{(8,9)G_{14}}$&$-_{(0,5)G_5}$\\[-1.4pt]
\jI{25.~Walk$(K_{ii})\x{,\, t=(\rho(A)+1)^{-1}}$                         }&$-_{\BiPe}     $&$-_{(0,5)G_5}$\\[-1.4pt]
\jI{26.~Estrada                                                          }&$-_{\BiPe}     $&$-_{(0,5)G_5}$\\[-1.4pt]
\jI{27.~Eigencentrality(Dice)\x{Dissimilarity)}                          }&$-_{(3,6)G_9}  $&$-_{(0,5)G_5}$\\[-1.4pt] 
\jI{28.~Eigencentrality(Jaccard)\x{Dissimilarity)}                       }&$-_{(3,6)G_9}  $&$-_{(0,5)G_5}$\\[-1.4pt] 
\jI{29.~\x{\ClosF}Closeness(Forest)\x{,\, $t=1$}                         }&$-_{(1,5)G_5}  $&$-_{(0,5)G_5}$\\[-1.4pt]
\jI{30.~\x{\ClosH}Closeness(Heat)\x{,\, $t=1$}                           }&$-_{(1,5)G_5}  $&$-_{(0,5)G_5}$\\[-1.4pt]
\jI{31.~\x{\ClolF}Closeness(logForest)\x{,\, $t=1$}                      }&$-_{(0,3)G_4}  $&$\!\!\!+/-_{(0,5)G_5}$\\[-1.4pt]
\jI{32.~\x{\ClolW}Closeness(logWalk)\x{,\, $t=(\rho(A)+1)^{-1}$}         }&$-_{(0,3)G_4}  $&$\!\!\!+/-_{(0,5)G_5}$\\[-1.4pt]
\jI{33.~\x{\ClolH}Closeness(logHeat)\x{,\, $t=1$}                        }&$-_{(0,3)G_4}  $&$-_{(0,5)G_5}$\\[-1.4pt]
\jI{34.~\x{\ClolC}Closeness(logComm.)\x{,\, $t=1$}             }&$-_{(0,3)G_4}  $&$-_{(0,5)G_5}$\\[-1.4pt]
\jI{35.~\x{\EcceF}Eccentricity(Forest)\x{,\, $t=1$}                      }&$-_{(0,1)G_7}  $&$-_{(0,5)G_5}$\\[-1.4pt]
\jI{36.~\x{\EcceH}Eccentricity(Heat)\x{,\, $t=1$}                        }&$-_{(0,1)G_7}  $&$-_{(0,5)G_5}$\\[-1.4pt]
\jI{37.~\x{\EcclF}Eccentricity(logForest)\x{,\, $t=1$}                   }&$-_{(0,1)G_7}  $&$-_{(0,5)G_5}$\\[-1.4pt]
\jI{38.~\x{\EcclW}Eccentricity(logWalk)\x{,\, $t=(\rho(A)+1)^{-1}$}      }&$-_{(0,1)G_7}  $&$-_{(0,5)G_5}$\\[-1.4pt]
\jI{39.~\x{\EcclH}Eccentricity(logHeat)\x{,\, $t=1$}                     }&$-_{(0,1)G_7}  $&$-_{(0,5)G_5}$\\[-1.4pt]
\jI{40.~\x{\EcclC}Eccentricity(logComm.)\x{,\, $t=1$}\!\!      }&$-_{(0,1)G_7}  $&$-_{(0,5)G_{5_{\mathstrut}}}$\\[-1.4pt]
\hline                      
\end{tabular}
} 
\end{table}

Table~\ref{ta_C-A} presents the results of verification of the Self-consistency and Bridge axioms for the centralities in~$\Ff$.
It turns out that only four measures\x{ in\x{ our set} $\Ff$\x{$=\{f_1\cdc f_{40}\}$}} satisfy Self-consistency. Four\x{!} other measures satisfy the Bridge axiom, including two measures that do so after a slight modification. When a measure violates an axiom, a record of the form `$-_{(u,v)G_k}\!$' indicates that the axiom is violated, in particular\x{among others}, for the ordered pair of nodes $(u,v)$ in graph~$G_k.$ Graphs $G_1$ to $G_{13}$ are shown in\x{familiar to us from} Fig.~\ref{f:SepGraphs}; two additional graphs found in Table~\ref{ta_C-A}, $G_{14}$ and $G_{15},$ are shown in Fig.~\ref{f:Florenc}. Graph $G'_5$\x{ $G_5\ssm\{\xz\{1,2\}\xz\}$ in this table} is \x{a\x{ shorthand} notation for the graph} obtained from $G_5$ by replacing $E(G_5)$ with $E(G_5)\ssm\{\xz\{1,2\}\xz\}.$

The following\x{ two} propositions\x{ present1/4} describe the positive results in Table~\ref{ta_C-A}.

\begin{proposition}[\hspace*{-.42em}\cite{Che23JCNCentr}]
The Generalized degree$,$\x{c} Eigenvector$,$\x{c} Katz$,$\x{c} and Bonacich\x{c} centralities satisfy Self-consistency.
\end{proposition}

The second result includes\x{involves measures} \ClolFa\ and \ClolWa\ introduced in {\em Materials and Methods}\x{Section~\ref{s:MatMet}}.
They are based on cutpoint additive distances as distinct from \ClolF\ and \ClolW\ included in $\Ff$.
This property enforces the Bridge axiom. 

\begin{proposition}[\hspace*{-.4em}\cite{Che23JCNCentr}]
The Closeness$,$\x{c} \ClolFa$,$ \ClolWa$,$ and Connectivity centralities satisfy the Bridge axiom.
\end{proposition}

It follows from Lemma~1 in \cite{Che23JCNCentr}\x{ Furthermore, one can show} that other strictly positive transitional measures \cite{Che11AAM} and cutpoint additive distances also produce\x{give rise to} centralities that satisfy the Bridge axiom.

\smallskip
Suppose the expert believes that all appropriate centralities \hla{must} satisfy Self-consistency (or Bridge axiom). This leads to a drastic reduction of the set $\Ff$\x{ of centralities} and\x{ to the corresponding reduction of} the culling survey.\x{ designed\x{ needed} to determine the most appropriate centrality measure.}

\begin{figure}[!t]
\begin{center}
\includegraphics[width=0.78\linewidth]{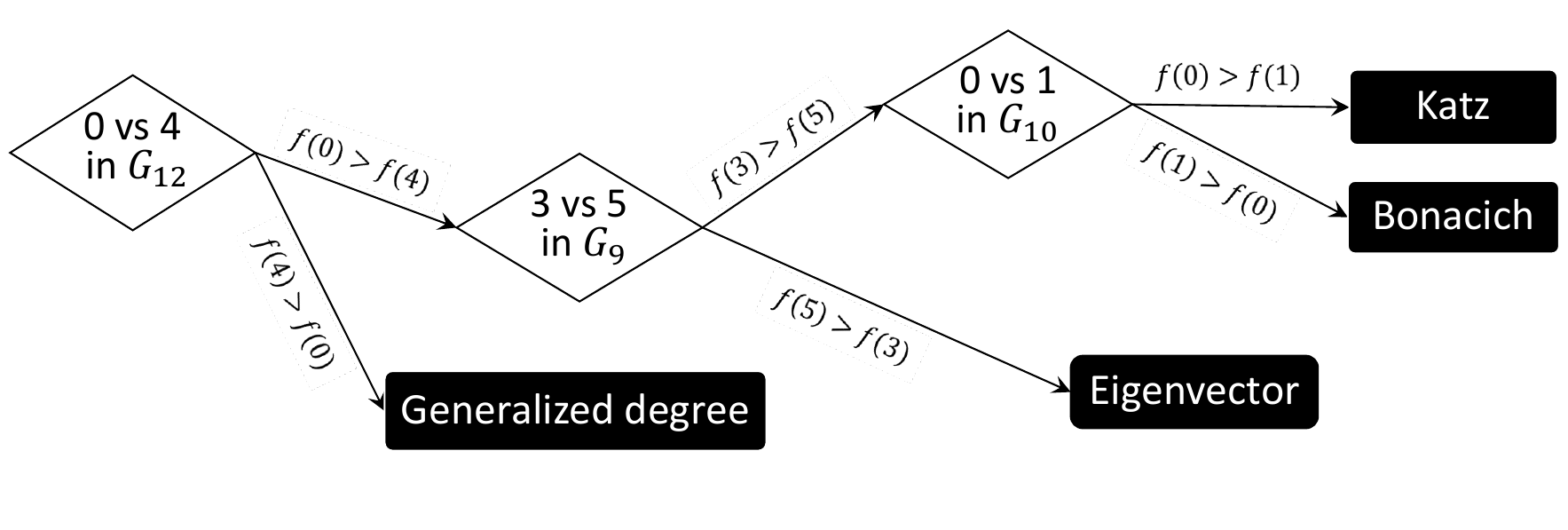} 

\Up{1.2}
\caption{A decision tree\x{survey} for the measures in $\Ff$ satisfying Self-consistency.\label{f:SC}}
\end{center}
\end{figure}

\begin{figure}[!t]
\begin{center}
\includegraphics[width=0.55\linewidth]{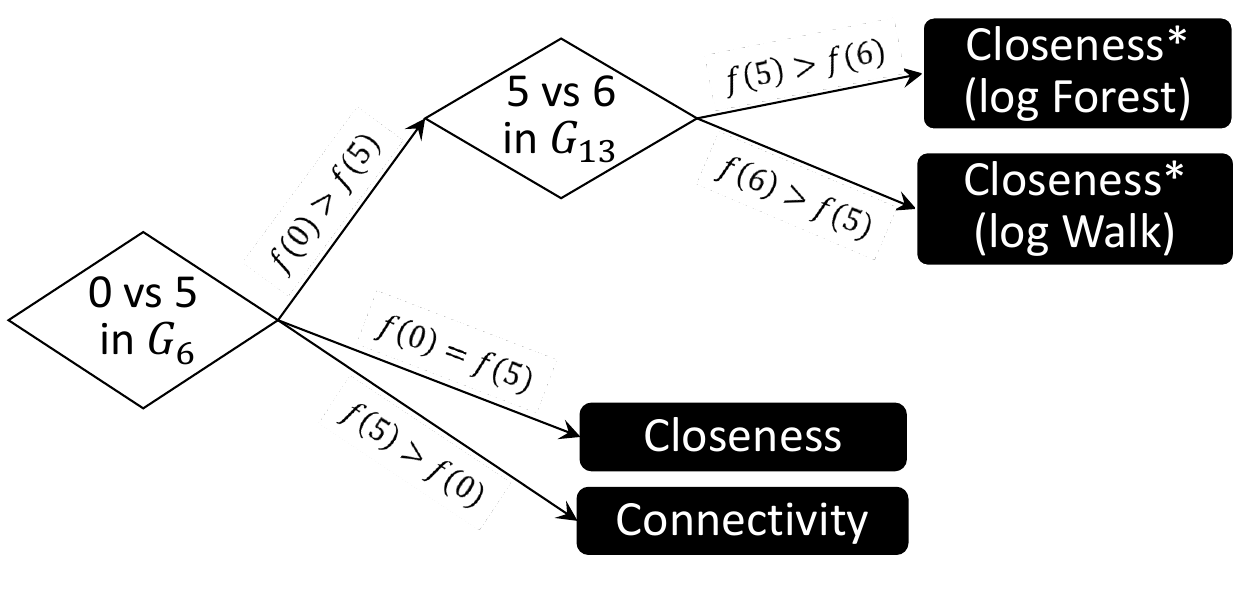} 

\Up{.3}
\caption{A decision tree\x{survey} for the four measures satisfying the Bridge axiom.\label{f:Bridge}
}
\end{center}
\end{figure}

The decision trees for the subsets of centralities satisfying the Self-Consistency or Bridge axiom are shown in
Fig.~\ref{f:SC} and Fig.~\ref{f:Bridge},
respectively.

The first one can be extracted from the general\x{ survey shown in} tree of Figures~\ref{f:Tree} and~\ref{f:STree}. The second\x{ survey} tree involves graphs that \separat{e} \ClolFa\ and \ClolWa\ from each other and the other measures that satisfy the Bridge axiom.

Some of the questions in the survey of\x{shown in} 
Fig.~\ref{f:SC} may not be easy to answer. This is due to the fact that\x{ because} the measures satisfying Self-consistency, as Lemma~2 in \cite{Che23JCNCentr} shows, have the same operating principle\x{framework, mechanism, spirit, are close relatives} based on \hla{supporting} neighborhood representations, so\x{ that} the difference between them can be quite subtle.
The comparison ``0 vs 1 in $G_{10}$'' can be replaced by ``2 vs 6 in $G_{11}$'' with $f(6)>f(2)$ leading to\x{ resulting in the} Katz centrality and $f(2)>f(6)$ yielding Bonacich centrality (which is\x{ just} a variation of Katz centrality)
with the\x{ chosen} parameter values specified in {\em Materials and Methods}\x{Section~\ref{s:MatMet}}.

\section{Discussion}
\label{s:Discuss}

\noindent In this paper, we proposed a method called culling for selecting a\x{ the most appropriate network} point centrality measure based on expert understanding of\x{beliefs about (opinion on)} how a measure suitable\x{appropriate} for a particular application should act\x{operate} on\x{ a set of simple} test graphs. The method consists of the following steps:
\begin{itemize}[itemsep=-.05em,topsep=.05em]
\item[(1)] Choosing\x{Forming} a finite set $\F$ of candidate measures, among which there are no rank equivalent measures;
\item[(2)] Generating a\x{ sequence} list of small (if possible) test\x{fairly (relatively)maximally the simplest simple, graphs as small as possible} graphs that \separat{e} all measures in~$\F$\x{ on some pairs of nodes};
\item[(3)] Compiling a decision-tree survey composed\x{consisting} of questions about the comparative centrality of test nodes in the test graphs;
\item[(4)] Completing the survey to obtain\x{\x{Since it is a decision tree,}This yields\x{provides}} a centrality measure consistent with\x{ all of} the expert's responses\x{preferences answers}.
\end{itemize}

The developed procedures\x{algorithms} \hlb{construct a decision-tree survey for any finite set $\F$ of rank-nonequivalent centrality measures. Indeed, even if measures $f_i$ and $f_j$ are rank-equivalent except for only one separating triple $(G,u,v),$ this triple will be found and included in the survey as the search procedure is complete.} 

\smallskip\smallskip
The conclusions are as follows.

1. Centrality measures based on rational ideas and satisfying reasonable axioms may behave counterintuitively (in the logic of a particular application or even in general) on small\x{simple} test\x{ low-density} graphs. This was seen in the examples of Weighted degree, PageRank, dissimilarity based Eigencentralities, and some other measures.\x{ in\x{ Introduction} the introduction and {\em Combination with the Axiomatic Approach\/}\x{(Sections~ \ref{s:intro} and~\ref{ss:Combining})}.} Therefore, examining the performance of measures on\x{ small} test graphs is a good complement\x{supplement} to typological, axiomatic, and \x{statistical}data-driven methods of analysis. Poor results of \x{these}such tests indicate inherent weaknesses\x{flaws} in\x{internal defects of} the measures and leave little chance\x{hope474/298} for \x{top}high-quality processing of real data.

2. The culling method relies on the expert's ability to judge with some certainty which of two nodes in a fairly simple graph has a higher centrality. We assume that this ability \x{is}can be based upon both the findings of prior studies 
and cumulative\x{accumulated} experience.
Moreover, the certainty of such a judgment is often stronger than the appeal of the heuristics behind a measure.
Given a set of centralities\x{y measures}, the culling method generates a decision-tree survey for selecting the measure that best fits \hla{the expert concept of centrality}\x{wishes}.
The implementation of the method varies depending on the\x{ preferred} choice of graphs for separating measures.
In this paper, we studied\x{considered} a test set $\Ff$ consisting of forty centralities, which\x{that} included, among others, several kernel-based closeness, eccentricity and other measures introduced in {\em Materials and Methods}\x{ Section~\ref{s:MatMet}}. It contained many\x{several} couples of closely related measures, and we were interested in whether large or dense graphs are necessary to separate them. The answer is no: we opted for $1$-trees (unicycles), and 13 graphs of this class of order $8$ (\hla{just}\x{only} one graph) or lower with node degrees at most $4$ were sufficient to separate all the measures in~$\Ff$.
For the case of an arbitrary set of measures, a general procedure was proposed.
%
%

3. If a set of measures $\F$ is updated\x{ resulting in a} to a new set $\F'\supset\F,$ then the culling method
adapts by extending the existing\x{ survey} decision tree to accommodate the changes.

4. The culling method generates a collection of\x{the separating} triples in the form of a graph and a pair of its nodes that distinguish each\x{ centrality} measure in the set $\F$ from\x{ the1/2.5} others. The centrality \x{ orders}\hla{inequalities/equalities} on these triples\x{These triples can be utilized to study the characteristics and peculiarities of the measures. The rankings\x{ offered} provided by a measure on the distinguishing graphs} form the ``profile'' of a measure characterizing its behavior and helping to decide whether to accept\x{approve1/12} or discard it.
They can also be used to \x{identify}\hla{classify} measures.
In this way, culling allows \x{us}one to reveal some\x{many unexpected} perplexing features of centralities. Examples have been mentioned in Item~1 of the conclusions. As another example consider the Bridging centrality~\cite{breitling2004rank\x{,hwang2008bridging}}. The expert\x{u-ser} who chooses it must\x{ accept} approve the following \hla{claims}\x{suggestions, judgements} of\x{ that} this measure\x{ sets} (see Fig.~\ref{f:SepGraphs}): $f(0)>f(2)$ in $G_2;$  $f(3)>f(4)$ in $G_4;$  $f(5)>f(3)$ in $G_6$ and $G_7;\,$ $f(5)>f(0)>f(1)$ in $G_8;$ $f(4)>f(2)$ in $G_9;$ $f(5)>f(1)$ in $G_{11},$ and $f(0)=f(1)$ in~$G_1.$

5. A decision tree for the entire set of known centralities\x{y measures} can be built by extending the\x{ decision} tree presented above (Figures~\ref{f:Tree} and~\ref{f:STree}).
Such a dendrogram may be useful (along with\x{in addition to} correlation-based methods~\cite{Saxena20CentrSurvey-}) for hierarchical clustering of measures.
This\x{ idea} is suggested by the fact that the\x{ decision} tree for $\Ff$ contains several subtrees whose leaves correspond to closely related centralities. For example, Callout~B (Fig.~\ref{f:STree}) consists of Eccentricity measures; Callout~C comprises Closeness measures and the related Decaying degree\x{ measure}; three measures satisfying Self-consistency belong to Callout~D, while the fourth\x{ measure} one is a ``sister'' of it.
The reason for this is that the\x{ culling} survey\x{ compiled using the procedure of Section~ \ref{ss:DecTree}} begins with the simplest graphs, while subtle differences between \hla{kindred}\x{similar, closely related} measures are distinguished by more complex graphs that appear later in\x{at the end of} the survey.

6. On the other hand,\x{However,} relying solely on culling for selecting a single output measure from an extensive set of input measures may not be entirely robust. 
Indeed, the corresponding survey may contain\x{, among others,} rather difficult questions, the answers to which may not be confident enough,
which can lead to an unstable result\x{outcome1/3}.
To enhance reliability, option ``not sure'' can be added to the survey forms, potentially leading to multiple outcome.
The\x{ final} choice among the resulting measures can be made by analyzing their application to real networks and checking their adherence to\x{compliance with1/2.7} credible axioms.
\x{Additionally}Moreover, the expert \hlb{can take the survey multiple times, each time with a distinct type of centrality in mind to get} a comprehensive picture using\x{utilizing relevant} centralities of several types (cf.~\cite{Vignery20Methodology,Brysbaert21Protein}). 

7. A promising application of the culling method is to\x{!} combine it with a normative approach by constructing decision trees for subsets of measures that satisfy the most credible axioms. The paper presents short surveys for the measures\x{ belonging to the set} in $\Ff$ (or their slight modifications) that meet the Self-consistency or Bridge axioms.
Thus, the recommended approach to address the question posed in the paper's title is to utilize a\x{ combined} strategy that encompasses several elements: verification of \hla{compliance with} the most important axioms, culling, consideration of computational complexity factors, experimental studies involving\x{ correlation analysis, Principal Component Analysis (PCA)} statistical methods and machine learning, as well as a typological perspective that connects network dynamics to the underlying assumptions and concepts of each centrality measure.

Among these components, the culling method stands out as one of the least labor-intensive and time-consuming. It relies on expert knowledge while its procedures are transparent.

\smallskip
This study identifies several directions for future research. Here are some of those\x{these} topics:
\begin{itemize}[itemsep=-.05em,topsep=.08em]%
\item Explore larger culling surveys;
\item Collect and analyze statistics of survey results from diverse experts to obtain\x{create} a measure attractiveness rating;
\item Construct decision trees for subsets of measures that satisfy crucial axioms, such as monotonicity conditions, and their combinations;
\item Adapt the culling method to centrality measures designed for directed networks;
\item Use variations in the construction of \x{culling}decision trees (cf. \cite{RokachMaimon15DMDTr})
    \hlb{to assess the robustness of the culling method}; 
\item Enhance the flexibility of culling surveys by \paa{a}~allowing the ``not sure'' option, \paa{b}~employing all graphs in $\G$ that \separat{e} specific pairs of measures, \paa{c}~allowing experts to utilize their own test graphs to select measures, \paa{d}~enabling experts to report confidence levels\x{+!} in their answers;
\item Create \hla{an interactive} web application implementing cul\-ling surveys for any chosen subset of centrality measures;
\item Apply the culling approach to the problem of selecting scoring methods for unbalanced tournaments\x{ mentioned in}~\cite{CheSha97a\x{,CheSha99}}.
\end{itemize}


%

\x{chen2018systematic,}
\x{other: fishburn1977condorcet, dequiedt2014local-previous version,}

\section{Materials and Methods}
\label{s:MatMet}
\subsection{Closeness and Eccentricity Centralities Induced by Graph Kernels} 
\label{ss:NewMes}

\noindent In this section, we present\x{introduce} several\x{some} new classes of centrality measures used\x{compared discussed} in {\em Test Survey\/} and subsequent sections\x{ of Results}. More information about them can be found in~\cite{avrachenkov2019similarities,Che23JCNCentr}.
\x{included in the test survey (see the section of the same name).}

Let $d(u,v)$ be the \emph{shortest path distance\/} 
between nodes $u$ and $v$\x{ in} of a graph $G$, i.e., the length of a shortest path between $u$ and~$v.$
Two popular 
distance based centrality measures are the [{shortest path}] \emph{Closeness\/} \cite{bavelas1948} 
\beq{e:clos} 
f(u)=\Big(\sum_{v\in V}d(u,v)\Big)^{-1},\quad u\in V
\eeq
and [{shortest path}] \Eccentricity\ \cite{bavelas1948} 
\beq{e:ecce} 
f(u)=(\max_{v\in V}d(u,v))^{-1},\quad u\in V.
\eeq

General classes of \Closeness\ and \Eccentricity\ \emph{centralities\/} are defined by \eqref{e:clos} and \eqref{e:ecce} with $d(u,v)$ being {\em arbitrary\/} distances for graph nodes.
In the literature, several classes of such distances and, more generally, dissimilarity measures have been proposed (see, e.g., \cite{Che13Paris,FoussSaerensShimbo16,avrachenkov2019similarities}). 
Many\x{Most} of\x{ the alternative such distances and dissimilarity measures} them are defined via graph kernels.
\x{We now}Let us consider four classes of them.

1. The parametric\x{ family of} {\it Katz\/} \cite{Katz53} {\it kernels\/} (also referred to as {\it Walk\x{ proximities}\/}~\cite{CheSha98} or {\it Neumann diffusion\/} \cite{FoussSaerensShimbo16} {\it kernels}) are defined as
\beq{e:PWalk}
P^{\Katz}(t) = \sum_{k=0}^\infty (tA)^k = (I-tA)^{-1}
\eeq
with $0<t<(\rho(A))^{-1},$ where $A=(a_{ij})$ is the adjacency matrix of $G$ and $\rho(A)$ is the spectral radius of~$A.$ 

2. The {\it Communicability kernels\/} \cite{estrada2005subgraph} are 
$$
P^{\Comm}(t) = \sum_{k=0}^\infty\frac{(tA)^k}{k!} = \exp(tA),\quad t>0.
$$

Two other\x{ families} classes\x{ of kernels} are defined similarly via the Laplacian matrix 
$$
L=\diag(A\bm1)-A, 
$$
where $\bm1=(1\cdc 1)^T$ and $\diag(\bm x)$ is the diagonal matrix with vector $\bm x$ on the diagonal.

3. The {\it Forest kernels}, or {\it regularized Laplacian kernels\/} \cite{CheSha95b} are 
\eq{e:PFor}{
P^{\For}(t) = (I + tL)^{-1},\;\; \mbox{where}\;\; t>0.
}

4. The {\it Heat kernels\/} are the Laplacian exponential diffusion kernels~\cite{KondorLafferty02diffusion}
$$
P^{\Hea}(t) = \sum_{k=0}^\infty\frac{(-tL)^k}{k!} = \exp(-tL),\quad t>0.
$$

By Schoenberg's theorem \cite{Schoenberg35\x{,Schoenberg38}}, if matrix $P=(p_{uv})$ is a kernel (i.e., is positive semidefinite), then it  produces a\x{generates a squared} Euclidean distance 
$d(u,v)$ by means of the transformation
\beq{e:K2D2} 
d(u,v)=\big(\tfrac12(p_{uu}+p_{vv}-p_{uv}-p_{vu})\big)^{\frac12},\quad u,v\in V,
\eeq
where $\frac12$ is the scaling factor.\x{ determines the scale. is related to scaling.}

Thus, all Walk, Communicability, Forest, and Heat kernels with appropriate parameter values provide distances\x{ $d(u,v)$} of the form \eqref{e:K2D2} whose substitution in\x{into5/10} \eqref{e:clos} and \eqref{e:ecce} yields\x{ [{\em generalized\/}]} {Closeness} and {Eccentricity} centralities. We will denote them by \Closeness(\xz{\em Kernel}\/) and {\Eccentricity(\xz{\em Kernel}\/)}\x{, respectively,} with the corresponding kernels substituted.

Furthermore, if $P_{n\times n}=(p_{uv})$ determines a {\it proximity measure\/} (\x{viz.,}which means that for any
$x,y,z\in V\x{ \{1\cdc n\}},$\; $p_{xy}+p_{xz}-p_{yz}\le p_{xx}$, and the inequality
is strict whenever $z=y$ and $y\ne x$), then \cite{CheSha98a} transformation\x{ \eqref{e:K2D2}}
\eq{e:K2D}{d(u,v)=\tfrac12(p_{uu}+p_{vv}-p_{uv}-p_{vu}),\quad u,v\in V}
provides\x{generates} a distance function that satisfies the axioms of a metric. The Forest kernel with any $t>0$ produces a proximity measure, while kernels in the remaining three classes\x{families} do so when $t$ is sufficiently small~\cite{avrachenkov2019similarities}. The centralities obtained from a {\em Proximity\/} measure\x{ through} by transformation \eqref{e:K2D} and substitution of the resulting distance into \eqref{e:clos} and \eqref{e:ecce} are denoted in \x{this}\hla{the present} paper by \Closeness$^*($\xz{\em Proximity}\/$)$ and \Eccentricity$^*($\xz{\em Proximity}\/$)$, respectively.

Moreover, if $P$ represents a strictly positive {\it transitional measure on~$G$\/} (i.e., $p_{xy}\,p_{yz}\le p_{xz}\,p_{yy}$ for all nodes $x, y,$ and $z,$ with $p_{xy}\,p_{yz} = p_{xz}\,p_{yy}$ whenever\x{iff} every path in $G$ from $x$ to $z$ visits $y$), then transformation
\beq{e:ln}
\hat{p}_{uv}=\ln p_{uv},\quad u,v\in V
\eeq
produces \cite{Che11AAM,Che13Paris} a proximity measure. In this case, \eqref{e:K2D} applied to $\hat{P}=(\hat{p}_{uv})$ reduces to
\eq{e:logdist}{d(u,v)=\tfrac12(\ln p_{uu}+\ln p_{vv}-\ln p_{uv}-\ln p_{vu})}
and generates \cite{Che13Paris} a {\it cutpoint additive distance\/} $d(u,v),$ viz., such a distance that $d(u,v) + d(v,w) = d(u,w)$ whenever\x{iff} $v$ is a cutpoint between $u$ and $w$ in $G$ (or, equivalently, whenever\x{iff} all paths connecting $u$ and $w$ visit~$v$).
The centralities obtained from any {\!\it Transitional\_Measure\/} by transformation \eqref{e:logdist} and substitution of the resulting distance into \eqref{e:clos} and \eqref{e:ecce}\x{ will be} are denoted by \Closeness$^*(${$\log$\em Transitional\_Measure}$)$ and \Eccentricity$^*(${$\log$\em Transitional\_Measure}$)$, respectively.

Since the Walk and Forest kernels determine \cite{Che11AAM} strictly positive transitional measures\x{ on the corresponding connected graph. Therefore}, transformation \eqref{e:logdist} applied to them generates cutpoint additive distances. Substituting them into \eqref{e:clos} and \eqref{e:ecce} produces {\em\ClolFa} \hla{and} {\em\ClolWa}, as well as the corresponding \Eccentricity$^*(\cdot)$\x{ centrality} measures.

Thus, based on the above properties, we define\x{{\it generalized}} {Closeness}\x{-type} and\x{{\it generalized}} {Eccentricity}\x{-type} centralities\x{ measures} obtained by substituting the:
\begin{itemize}[itemsep=-.05em,topsep=.085em] 
\item Forest kernel;
\item Heat kernel;
\item logarithmic Forest kernel;
\item logarithmic Walk kernel;
\item logarithmic Heat kernel, and
\item logarithmic Communicability kernel
\end{itemize}
%
transformed by \eqref{e:K2D2} or \eqref{e:K2D} into \eqref{e:clos} and~\eqref{e:ecce}.
These centralities are included in the test survey\x{ (discussed) presented below}.
We set $t=1$ for the Forest, Heat, and Communicability kernels and $t=(\rho(A)+1)^{-1}$ for the Walk kernel.
Regarding other measures\x{Furthermore}, we use PageRank centrality with $\alpha=0.85$, Generalized degree with $\ve=2$, Decay centrality with $\delta=0.9$, Katz centrality with $a=(\rho(A)+1)^{-1}$, and Bonacich centrality (which is a variation of Katz centrality) with $\alpha=1$ and $\beta=(\rho(A)+0.5)^{-1}.$ 
The difference between the Katz and Bonacich centralities\x{last two measures is determined by} \x{reduces to}\hla{lies in} the difference \x{of}\hla{in} their parameters $a$ and~$\beta$.

The~\Closeness$^*$(\xz\Forest) centrality\x{ (without square-rooting)} was examined in \cite{jin2019forest}\x{, where the  authors conclude} with the conclusion that ``forest distance centrality has a better discriminating power than alternate metrics such as betweenness, harmonic centrality, eigenvector centrality, and PageRank.''
The authors of \cite{jin2019forest} believe\x{claim note add} that the order of node importance\x{ given} induced by the forest distances on\x{ some certain} simple graphs is ``\x{ is in agreement with their}consistent with human intuition.''
\x{C}The $t\to\infty$ case of this centrality is \cite{CheSha01} the Resistance centrality\x{, which features} featuring in~\cite{Tyloo19oscill}.

\subsection{\x{Some }Other Kernel-based Centralities} 
\label{ss:OthMes}

\noindent \x{It should be emphasized that}While the above measures are meaningful\x{ promising noteworthy important(can be considered as) examples of} kernel-based centralities,\x{ there are other} they do not exhaust all kernels and transformations \cite{FoussSaerensShimbo16,avrachenkov2019similarities} that can be used to obtain such measures.
To mention some\x{ other approaches} alternative constructions, note that \x{every}each distance on graph nodes 
can be integrated in the harmonic closeness framework\x{la} \cite{marchiori2000harmony}, $p$-Means framework~\cite{deAndrade2019p}, or the framework developed in~\cite{agneessens2017geodesic}.

\x{ Moreover,(Finally,)In addition to the above approaches}Furthermore, kernels and similarity/proximity measures can\x{ produce} be used to obtain centralities directly, without transformations into distances. \x{One of the measures of this kind}An example of such measures is the \EstradaSubgraph\ {\em centrality}~\cite{estrada2005subgraph}. This index\x{measure} of a graph node $u$ is equal to the diagonal entry $p^{\Comm}_{uu}$ of the Communicability kernel, so we denote it by \Communicability$(K_{ii}).$
Similarly, \Walk$(K_{ii})$ is the measure $f(u)=p^{\Katz}_{uu},$ $u\in V$ determined by the diagonal entries of the Walk kernel.

\x{Consider now }One more type of centrality measures\x{ of (a different) another type are, which are} is constructed by summing the non-diagonal entries of the rows of a kernel matrix. We consider the measures of this kind \Communicability($K_{ij}$) and \Walk($K_{ij})$ defined by $f(u)=\sum_{v\ne u}p^{\Comm}_{uv}$ and $f(u)=\sum_{v\ne u}p^{\Katz}_{uv},$ $u\in V,$ 
respectively. Finally,\x{Furthermore,} \TotalCommunicability\ \cite{benzi2013total} is obtained by summing {\em all\/} row entries of the Communicability kernel: $f(u)=\sum_{v\in V}p^{\Comm}_{uv}$; it can be described \cite{deMeoProvetti2019general} in terms of ``potential gain.'' \x{similarly to (as well as) the corresponding \Katzc\ measure, which}
The \TotalWalk\ measure $f(u)=\sum_{v\in V_{\mathstrut}}p^\Katz_{uv}=(P^\Katz\xy\bm1)_u$ is order equivalent to the \KatzcCentrality\ \cite{Katz53} expressed as\x{{by57/121}} $f(u)=((P^\Katz-I)\bm1)_u$. 

Note that the existence of hundreds of types and subtypes of centrality, coupled by the existence of infinite families of measures, highlights the need for powerful tools for comparing\x{ and discriminating between} centrality indices and selecting\x{choosing} the most appropriate ones.
%
%
%

\section*{Acknowledgments}

\noindent The authors thank Anna Khmelnitskaya and Konstantin Avrachenkov for helpful\x{ valuable} discussions and Matthew Jackson for an important remark at the final stage.\x{ conversations.} \hlb{We are grateful to the anonymous referees for their \aB{valuable} comments.}
%
%
%
%

\bibliographystyle{IEEEtran} 
\bibliography{IEEEabrv,centrality-TSMC}           

\end{document}